\documentclass{article}
\usepackage[T1]{fontenc}
\usepackage[utf8]{inputenc}
\usepackage{graphicx}
\usepackage{subfigure}
\usepackage{authblk}
\usepackage{url}
\usepackage[version=3]{mhchem}
\usepackage[detect-all]{siunitx}
\DeclareSIUnit{\Molar}{\textsc{m}}
\usepackage{amsmath}
\usepackage{amssymb}
\usepackage{amsthm}
\usepackage{amsfonts}
\usepackage{xcolor}
\usepackage{microtype}

\usepackage{anysize}
\marginsize{2cm}{2cm}{.3cm}{2cm}

\title{Thermal switch of oscillation frequency \\
in Belousov-Zhabotinsky liquid marbles}
\author[1]{Andrew Adamatzky\footnote{Contact: andrew.adamatzky@uwe.ac.uk}}
\author[1]{Claire Fullarton}
\author[1]{Neil Phillips}
\author[1,2]{Ben De~Lacy~Costello}
\author[1]{Thomas C.\ Draper}
\affil[1]{Unconventional Computing Laboratory, Department of Computer Science and Creative Technologies, University of the West of England, Bristol, BS16 QY, UK}
\affil[2]{Institute of Biosensing Technology, Centre for Research in Biosciences, University of the West of England, Bristol, BS16 1QY, UK}

\date{}

\graphicspath{{figs/}}

\begin{document}

\maketitle

\begin{abstract}
\noindent
External control of oscillation dynamics in the Belousov-Zhabotinsky reaction is important for many applications including encoding computing schemes. When considering the BZ reaction there are limited studies dealing with thermal cycling, particularly cooling, for external control. Recently, liquid marbles (LMs) have been demonstrated as a means of confining the Belousov-Zhabotinsky reaction in a system containing a solid-liquid interface. BZ LMs were prepared by rolling \SI{50}{\micro\litre} droplets in polyethylene (PE) powder. Oscillations of electrical potential differences within the marble were recorded by inserting a pair of electrodes through the LM powder coating into the BZ solution core. Electrical potential differences of up to \SI{100}{\mV} were observed with an average period of oscillation ca. \SI{44}{\second}. BZ LMs were subsequently frozen to -1\textsuperscript{o}C to observe changes in the frequency of electrical potential oscillations. The frequency of oscillations reduced upon freezing to \SI{11}{\milli\Hz} cf.\ \SI{23}{\milli\Hz} at ambient temperature. The oscillation frequency of the frozen BZ LM returned to \SI{23}{\milli\Hz} upon warming to ambient temperature. Several cycles of frequency fluctuations were able to be achieved.

\vspace{0.1cm}

\end{abstract}

\section{Introduction}

Space-time dynamics of oxidation wave-fronts, including target waves, spiral waves, localised wave-fragments and combinations of these, in a non-stirred Belousov-Zhabotinsky (BZ) medium~\cite{belousov1959periodic, zhabotinsky1964periodic} have been used to implement information processing since seminal papers by Kuhnert, Agladze and Krinsky~\cite{kuhnert1986new, kuhnert1989image}.  The spectrum of unconventional computing devices prototyped with BZ reaction is rich. Examples include image processing and memory~\cite{kaminaga2006reaction},  diodes~\cite{DBLP:journals/ijuc/IgarashiG11}, geometrically constrained logical gates~\cite{steinbock1996chemical}, controllers for robots~\cite{adamatzky2004experimental},  wave-based counters~\cite{gorecki2003chemical}, neuromorphic architectures~\cite{gorecki2006information,  gentili2012belousov,gruenert2015understanding, stovold2017associative},  and binary arithmetical circuits~\cite{adamatzky2010slime, suncrossover,stevens2012time}. 

While most of BZ computing devices use the presence of a wave-front in a selected locus of space as a manifestation of logical {\sc True}, there is a body of works on information coding with frequencies of oscillations. Thus, Gorecki et al~\cite{gorecki2014information} proposed to encode {\sc True} as high frequency and {\sc False} as low frequency: {\sc or} gates, {\sc not} gates and a diode have been realised in numerical models. Other results in BZ frequency based information processing include  frequency transformation with a passive barrier~\cite{sielewiesiuk2002passive}, frequency band filter~\cite{gorecka2003t}, and memory~\cite{gizynski2017chemical}.  Using frequencies is in line with current developments in oscillatory logic~\cite{borresen2012oscillatory}, fuzzy logic~\cite{gentili2012belousov}, oscillatory associated memory~\cite{calayir2013fully}, and computing in arrays of coupled oscillators~\cite{baldi1990computing,nikonov2015coupled}. Therefore, frequencies of oscillations in BZ media will be the focus of this paper.

%Therefore, in the present paper we focus on frequency of oscillations in the BZ medium. 

Most prototypes of BZ computers involve some kind of geometrical constraining of the reaction: a computation requires a compartmentalisation. An efficient way to compartmentalise BZ medium is to encapsulate it in a lipid membrane~\cite{jones2015autonomous,tomasi2014chemical}. This encapsulation enables the arrangement of elementary computing units into elaborate computing circuits and massive-parallel information processing arrays~\cite{adamatzky2012architectures,holley2011logical,adamatzky2011polymorphic,gizynski2017evolutionary}. BZ vesicles have a lipid membrane and therefore have to reside in a solution phase, typically oil, and they are susceptible to disruption of the lipid vesicles through natural ageing and mechanical damage. Thus potential application domains of the BZ vesicles are limited. This is why in the present paper we focus on liquid marbles, which offer  us capability for `dry manipulation' of the compartmentalized oscillatory medium. BZ-LMs also provide the possibility for active transport processes which is not easily possible with vesicles. The liquid marbles (LM), proposed by  Aussillous and Qu\'{e}r\'{e} in 2001~\cite{aussillous2001liquid}, are liquid droplets coated by hydrophobic particles at the liquid/air interface. The liquid marbles do not wet surface and therefore can be manipulated by a variety of means~\cite{ooi2015manipulation}, including rolling, mechanical lifting and dropping, sliding and floating~\cite{bormashenko2009mechanism,draper2018a,celestini2018propulsion}. A range of applications of LMs is huge and spans most field of life sciences, chemistry, physics and engineering~\cite{bormashenko2011liquid,mchale2011liquid,draper2017,avruamescu2018liquid,daeneke2018liquid}. Recently we demonstrated that the BZ reaction is compatible with typical LM chemistry: BZ-LMs support localised excitation waves, and non-trivial patterns of oscillations are evidenced in ensembles of the BZ LMs~\cite{fullarton2018belousov}. 

Oscillations in the BZ reaction media can be controlled by varying the concentrations of chemical species involved in the reaction, and with light~\cite{petrov1993controlling,kadar1997reaction}, mechanical strain~\cite{yashin2009controlling}, and temperature~\cite{blandamer1975investigation,vajda1988cryo,masia2001effect,ito2003temperature,bansagi2009high}. While a number of high impact results on the thermal sensitivity have been published, the topic still remains open and of utmost interest. Moreover, in LMs we might have difficulties in controlling the reaction with illumination because most types of hydrophobic coating are not perfectly transparent and absorb wavelengths of light important for exerting control over the BZ reaction.  This is why in the present manuscript we focus on thermal control and tuning of the oscillations.

Temperature sensitivity of the BZ reaction was initially substantially analysed by Blandamer and Morris~\cite{blandamer1975investigation} who, in 1975, showed a dependence of the frequency of oscillations of a redox potential in a stirred BZ reaction with a change in temperature. Periods of oscillations reported were 190~s at 25\textsuperscript{o}C, 70~s at 35\textsuperscript{o}C, and 40~s at 45\textsuperscript{o}C. In 1988 Vajda et al.~\cite{vajda1988cryo} demonstrated that temporal oscillations of a BZ mixture persist in a frozen aqueous solution at -10\textsuperscript{o}C to -15\textsuperscript{o}C. By tracing  \ce{Mn^{2+}} ion signal amplitude they showed that the frozen BZ solutions oscillate 3 times, at -10\textsuperscript{o}C, and 11 times, at -15\textsuperscript{o}C, faster than liquid phase BZ. The oscillation frequency increase has been explained by formation of crystals and interfacial phenomena during freezing. This might be partly supported by experiments with chlorite-thiosulphate system frozen to -34\textsuperscript{o}C~\cite{szirovicza1991propagating}. There a velocity of wave fronts is increased because en route to total freezing the reaction occurs only in the thin liquid layer, at the periphery of the solid domain, where concentrations of chemicals are temporarily higher. In 2001 Masia at al.~\cite{masia2001effect} monitored oscillations in non-stirred BZ in a batch reactor of 4~\ce{cm^3} by the solution absorbency at 320~nm. The reactor was kept at various temperatures through thermostatic control. They reported periodic oscillation at temperatures  0\textsuperscript{o}C to 3\textsuperscript{o}C, quasi-periodic at 4\textsuperscript{o}C to 6\textsuperscript{o}C, and chaotic at 7\textsuperscript{o}C to 8\textsuperscript{o}C. Bansagi et al.~\cite{bansagi2009high} experimentally demonstrated that by increasing temperature from 40\textsuperscript{o}C to 80\textsuperscript{o}C it is possible to obtain oscillations of frequency over 10Hz; they also showed that the frequency of oscillations grows proportionally to temperature (in the range studied). Ito et al.~\cite{ito2003temperature} reported linear dependence of an oscillation period --- of polymers impregnated with BZ --- from temperature in the range 5~\textsuperscript{o}C to 25~\textsuperscript{o}C.

We establish an electrical interface with BZ LMs by piercing them with a pair of electrodes. This is done for two reasons. First, the coating of LMs is usually non-transparent therefore conventional optical means of recording oxidation dynamics would not be sufficient. In addition marbles are 3D structures and there is evidence that they support complex 3D waves, therefore, electrodes positioned within the marble potentially allow the 3D oscillation dynamics to be mapped whereas imaging is difficult to interpret from a 3D standpoint. Second, our ultimate goal is to implement an unconventional computing device with BZ LMs. Such devices rarely stand-alone but are usually interfaced with conventional electronics, thus electrical recording seemed to be most appropriate.

\section{Methods}
\label{methods}

Belousov-Zhabotinsky (BZ) liquid marbles (LMs) were produced by coating droplets of BZ solution with ultra high density polyethylene (PE) powder (Sigma Aldrich, CAS 9002-88-4, Product Code 1002018483, particle size $150 \mu m$). The BZ solution was prepared using the method reported by Field~\cite{Field1979}, omitting the surfactant Triton X. \SI{18}{\Molar} Sulphuric acid \ce{H2SO4} (Fischer Scientific), sodium bromate \ce{NaBrO3}, malonic acid \ce{CH2(COOH)2}, sodium bromide \ce{NaBr} and \SI{0.025}{\Molar} ferroin indicator (Sigma Aldrich) were used as received. Sulphuric acid (\SI{2}{\ml}) was added to deionised water (\SI{67}{\ml}), to produce \SI{0.5}{\Molar} \ce{H2SO4}, \ce{NaBrO3} (5g) was added to the acid to yield \SI{70}{\ml} of stock solution (0.48M).

Stock solutions of \SI{1}{\Molar} malonic acid and \SI{1}{\Molar} \ce{NaBr} were prepared by dissolving \SI{1}{\gram} in \SI{10}{\ml} of deionised water. In a \SI{50}{\ml} beaker, \SI{0.5}{\ml} of \SI{1}{\Molar} malonic acid was added to \SI{3}{\ml} of the acidic \ce{NaBrO3} solution. \SI{0.25}{\ml} of \SI{1}{\Molar} \ce{NaBr} was then added to the beaker, which produced bromine. The solution was set aside until it was clear and colourless (ca.\ \SI{3}{\minute}) before adding \SI{0.5}{\ml} of \SI{0.025}{\Molar} ferroin indicator.

BZ LMs were prepared by pipetting a 75\ce{\mu}L droplet of BZ solution, from a height of ca.\ \SI{2}{\mm} onto a powder bed of PE, using a method reported previously \cite{fullarton2018belousov}. The BZ droplet was rolled on the powder bed for ca.\ \SI{10}{\second} until it was fully coated with powder.

\begin{figure}[!tbp]
    \centering
   \subfigure[]{\includegraphics[width=0.25\textwidth]{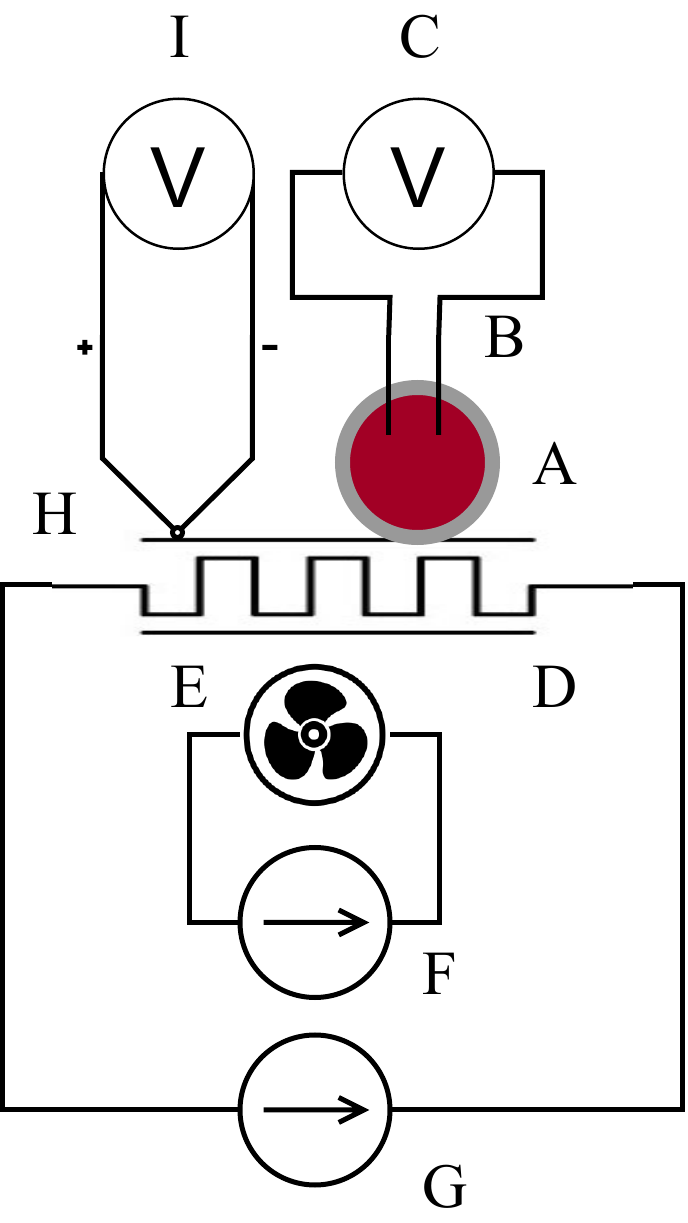}\label{schemesetup}}
    \subfigure[]{\includegraphics[width=0.6\textwidth]{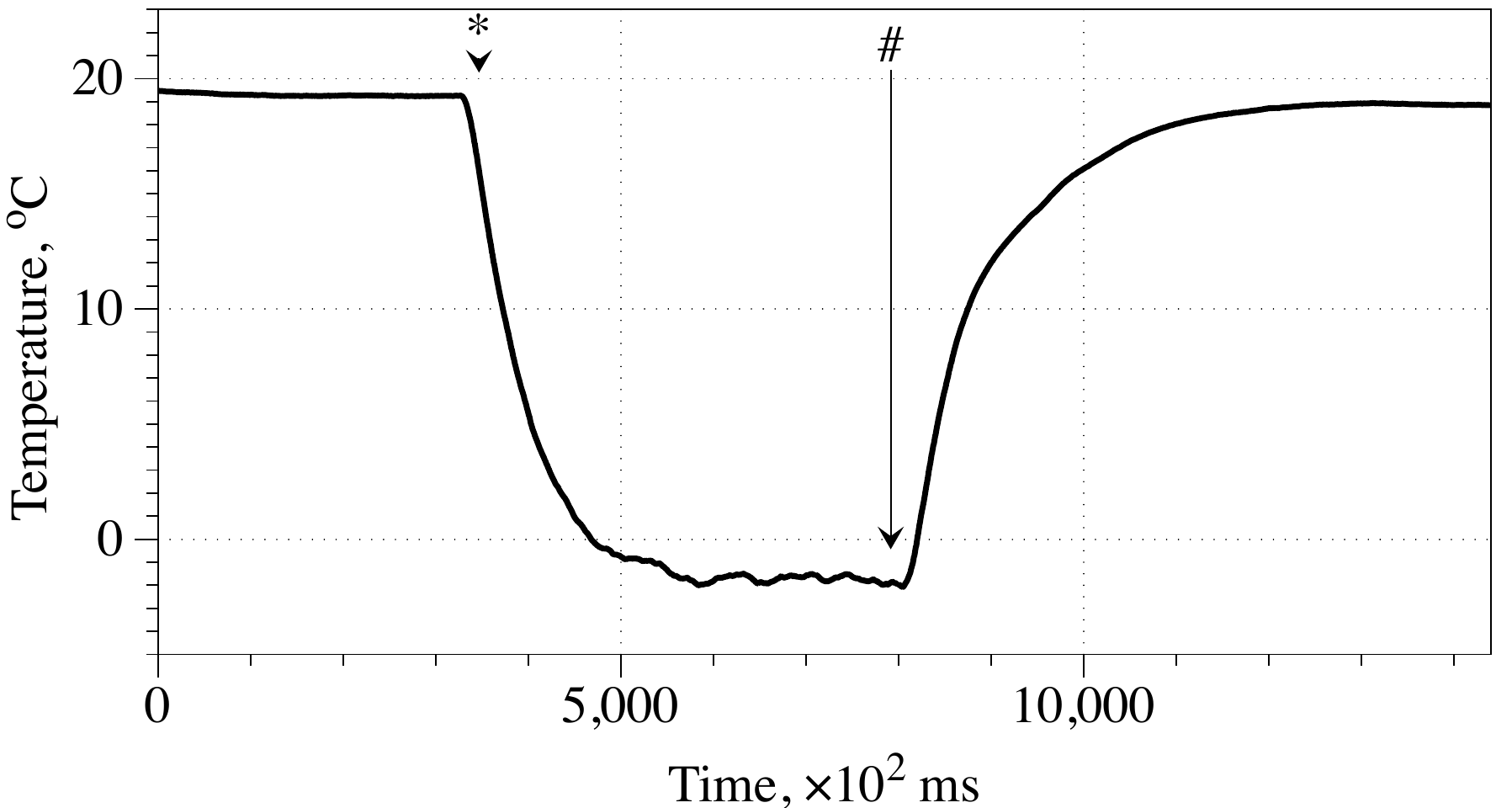}\label{temperature}}
    \caption{Experimental setup.
    (a)~A scheme of the setup: A -- BZ LM, B -- a pair of electrodes, C --- Pico ADC-24 logger, D -- Peltier element, E -- fans, F -- power supply for fans, G -- power supply for the Peltier element, H -- thermocouple, I -- TC-08 thermocouple data logger.
    (b)~Dynamics of temperature on the surface of the Petri dish when the Peltier element is powered by 7V. The moment of power on is shown by `*' and switched off, `\#'.}
    \label{fig:setup}
\end{figure}

A scheme of experimental setup is shown in Fig.~\ref{schemesetup}. 
A LM was placed in Petri dish (35 mm diameter) and pierced with two iridium coated stainless steel electrodes (sub-dermal needle electrodes with twisted cables (SPES MEDICA SRL Via Buccari 21 16153 Genova, Italy). Electrical potential difference between electrodes was recorded with a Pico ADC-24 high resolution data logger (Pico Technology, St Neots, Cambridgeshire, UK), sampling rate  25~ms.

A Petri dish with LM was mounted to a Peltier element (100~W, 8.5~A, 20~V, 40$\times$40~mm,  RS Components Ltd., UK ), which in turn was fixed to an aluminium heat sink, with Silver CPU Thermal Compound,  cooled by two 12V fans (powered separately from the Peltier element). Temperature at the Peltier element was controlled via RS PRO Bench Power Supply Digital (RS Components Ltd.
Birchington Road, Corby, Northants, NN17 9RS, UK).  Temperature at the bottom of the Petri dish was monitored using TC-08 thermocouple data logger (Pico Technology, St Neots, Cambridgeshire, UK), sampling rate 100~ms. A typical cooling rate was -1\ce{^o}C per 10~s, and warming rate +1\ce{^o}C per 20~s, exact shape of the functions is shown in Fig.~\ref{temperature}. 

\section{Results}

\begin{figure}[!tbp]
    \centering
    \subfigure[]{\includegraphics[width=0.15\textwidth]{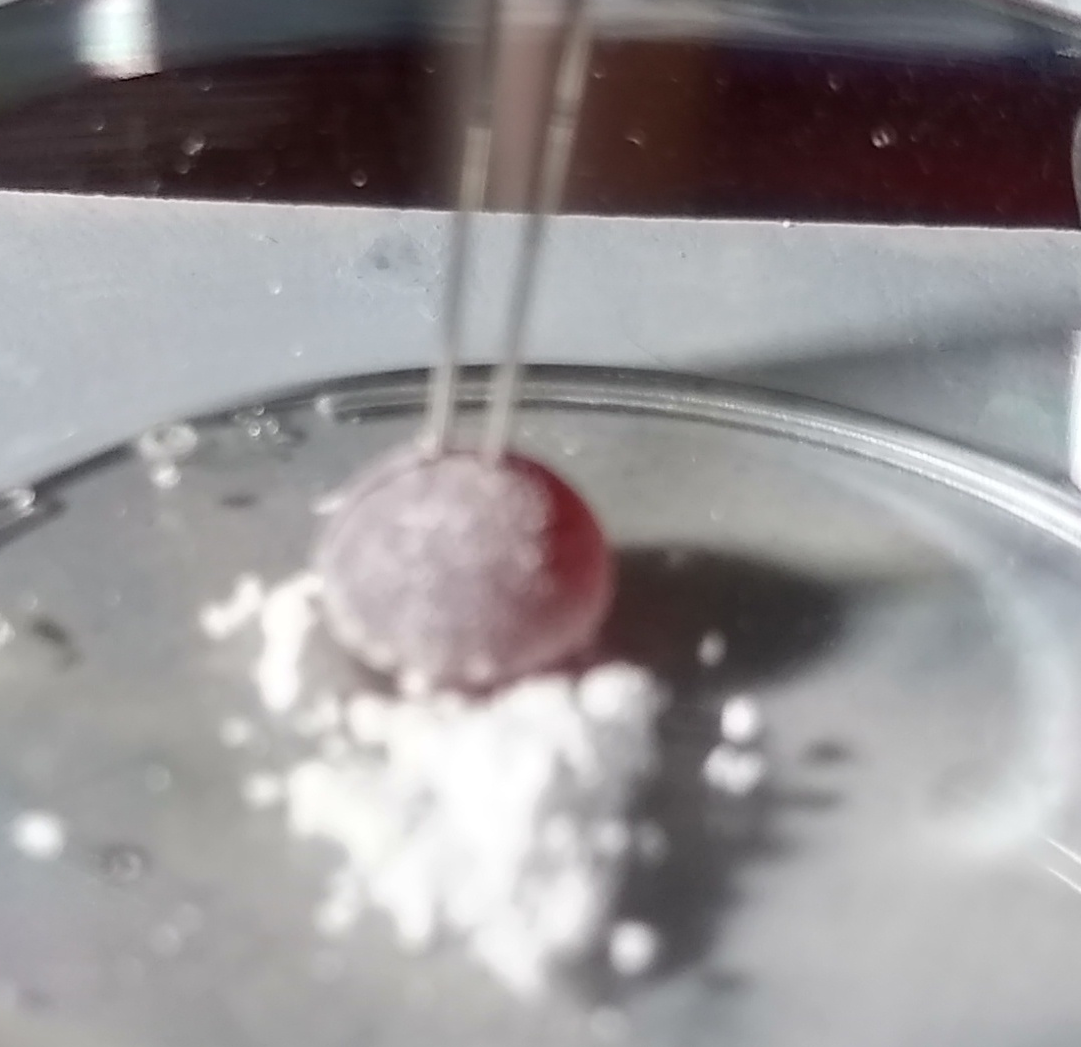}\label{intactLM1}}
    \subfigure[]{\includegraphics[width=0.16\textwidth]{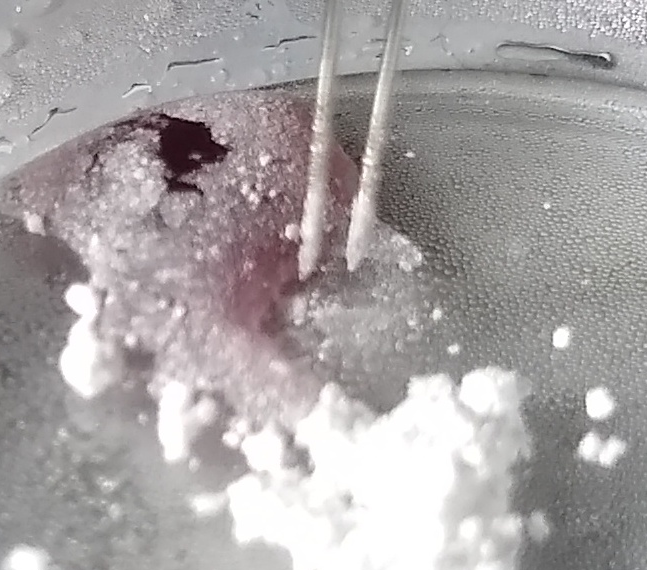}\label{burstLM1}}
      \subfigure[]{\includegraphics[width=0.6\textwidth]{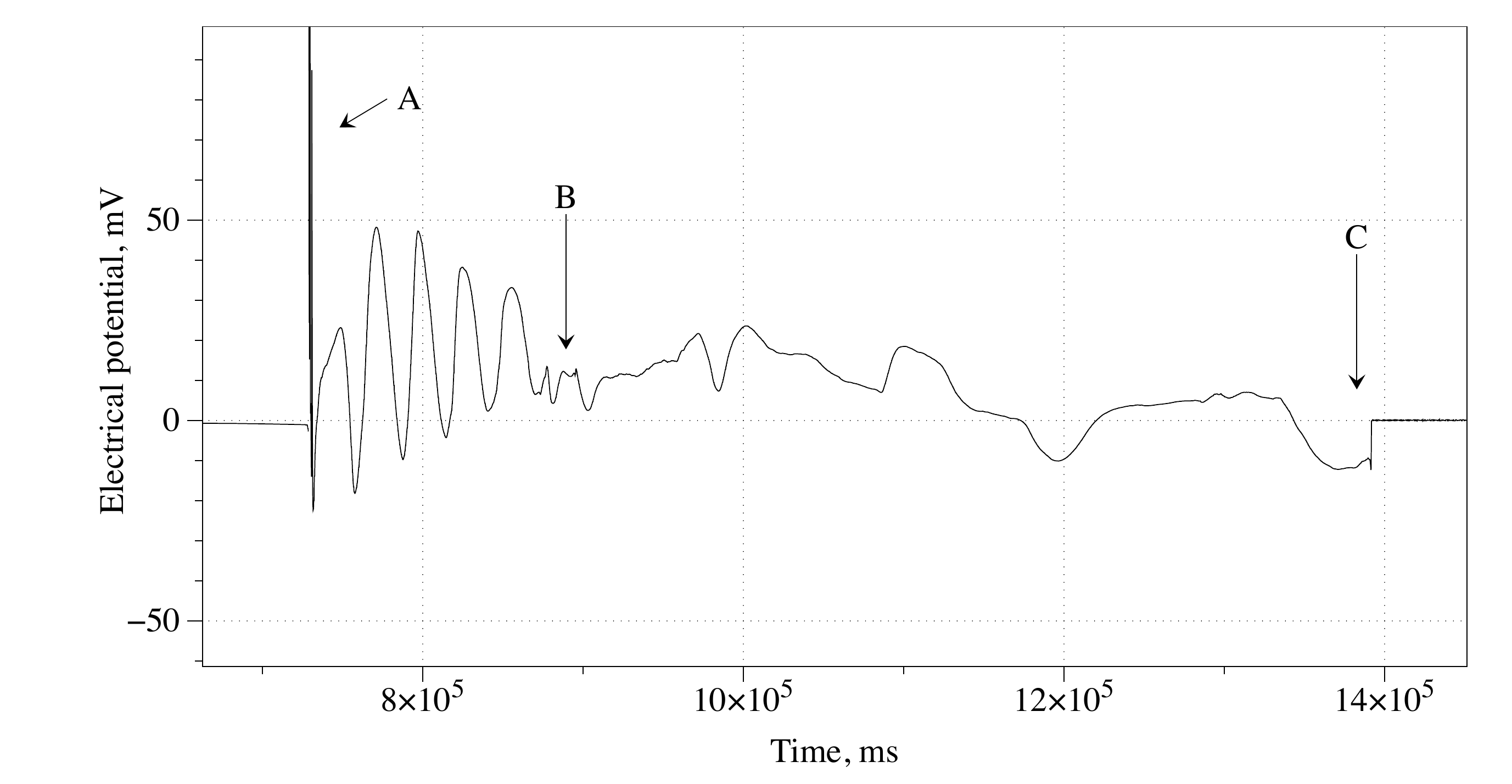}\label{plot1}}
     \subfigure[]{\includegraphics[scale=0.09]{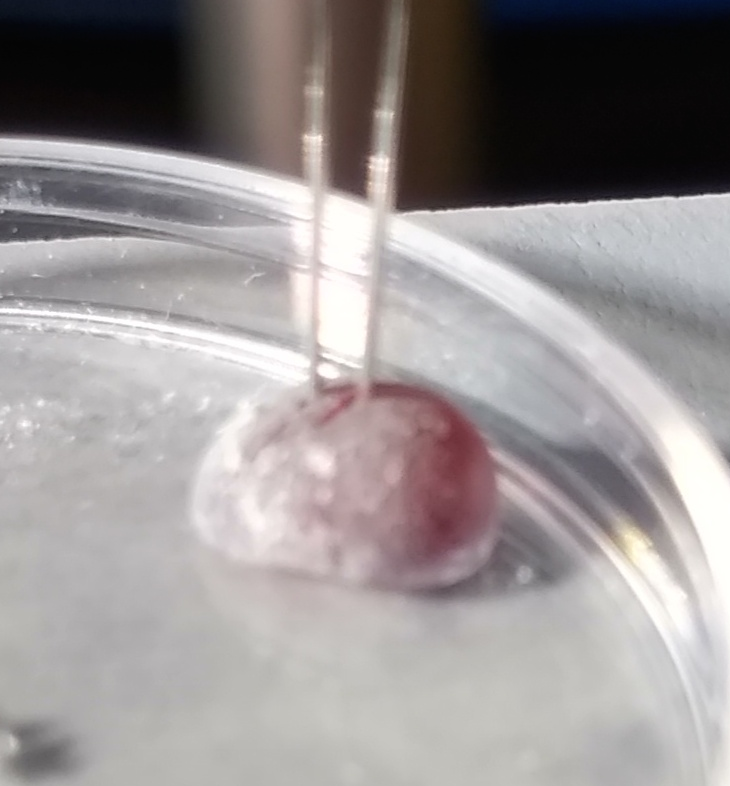}\label{intactLM2}}
      \subfigure[]{\includegraphics[scale=0.1]{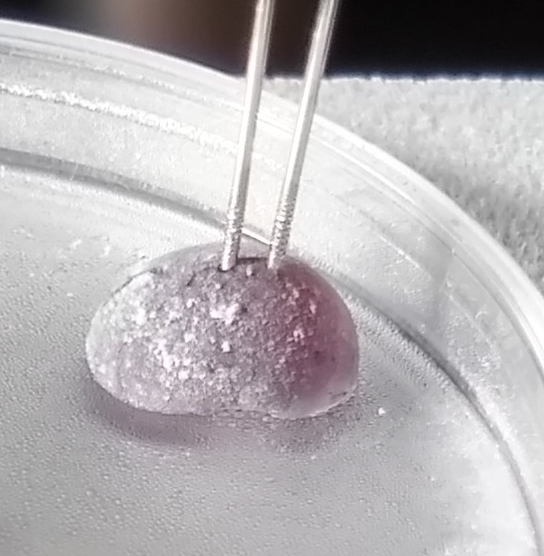}\label{FreezingLM2}}
    \subfigure[]{\includegraphics[scale=0.1]{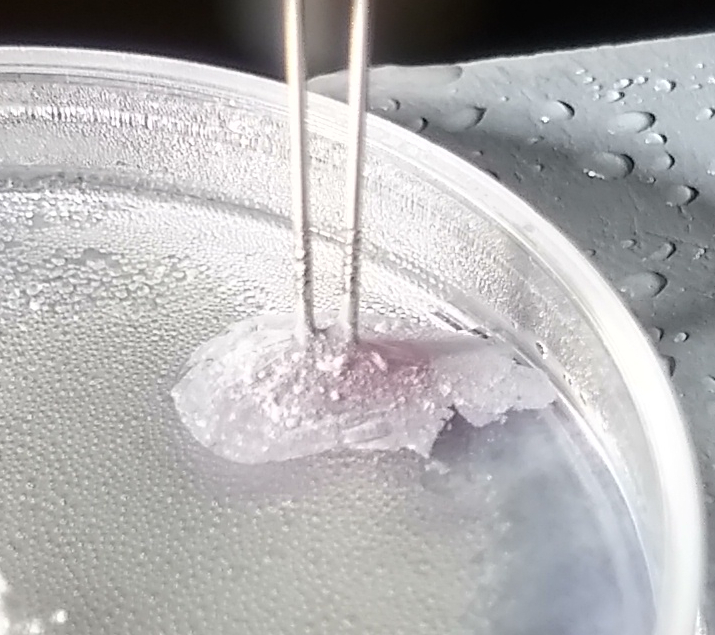}\label{BurstLM2}}
    \subfigure[]{\includegraphics[width=0.6\textwidth]{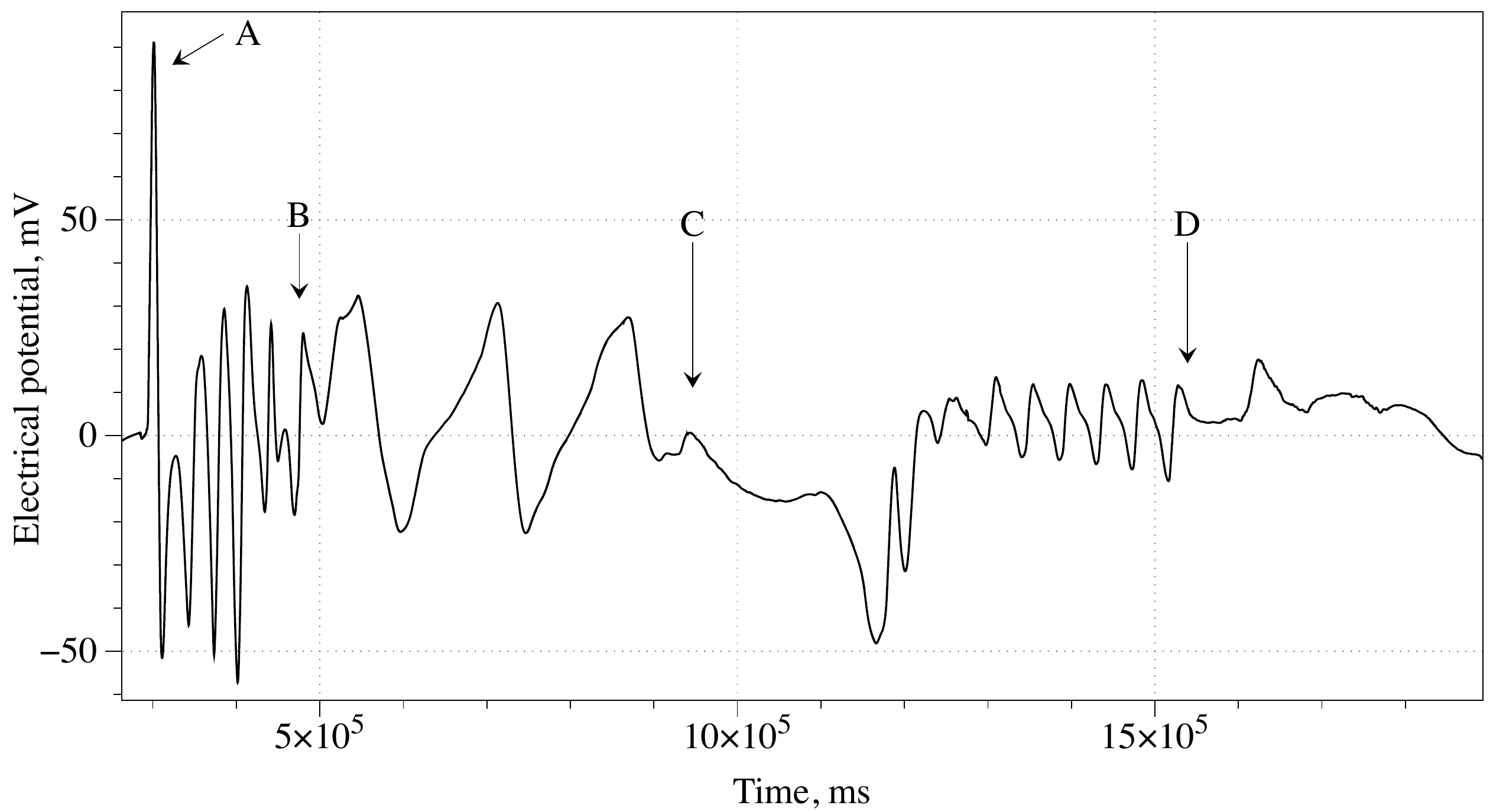}\label{plot2}}
    \caption{(a--c)~LM bursts at first freezing. (d--g)~LM burst at second freezing. 
    (a)~Marble at the beginning of experiment, (b) Marble burst at some point of freezing. (c) plot of oscillations: A -- the marble is stimulated with a silver wire, B -- oscillations started, Peltier element is switched on, C -- marbles cools downs, eventually the marbles bursts.
    (d)~LM at the beginning of experiment. (e)~Cooled down LM, 
    (f)~LM bursts and spreads at the second round of freezing.
    (g)~Dynamics of electrical potential:
    A -- marble is stimulated by a silver wire for 2-3~sec, B -- Peltier element is switched on, C -- Peltier is switched off, D -- Peltier is switched on again.}
    \label{fig:burst}
\end{figure}

\begin{figure}[!tbp]
    \centering
     \subfigure[]{\includegraphics[width=0.49\textwidth]{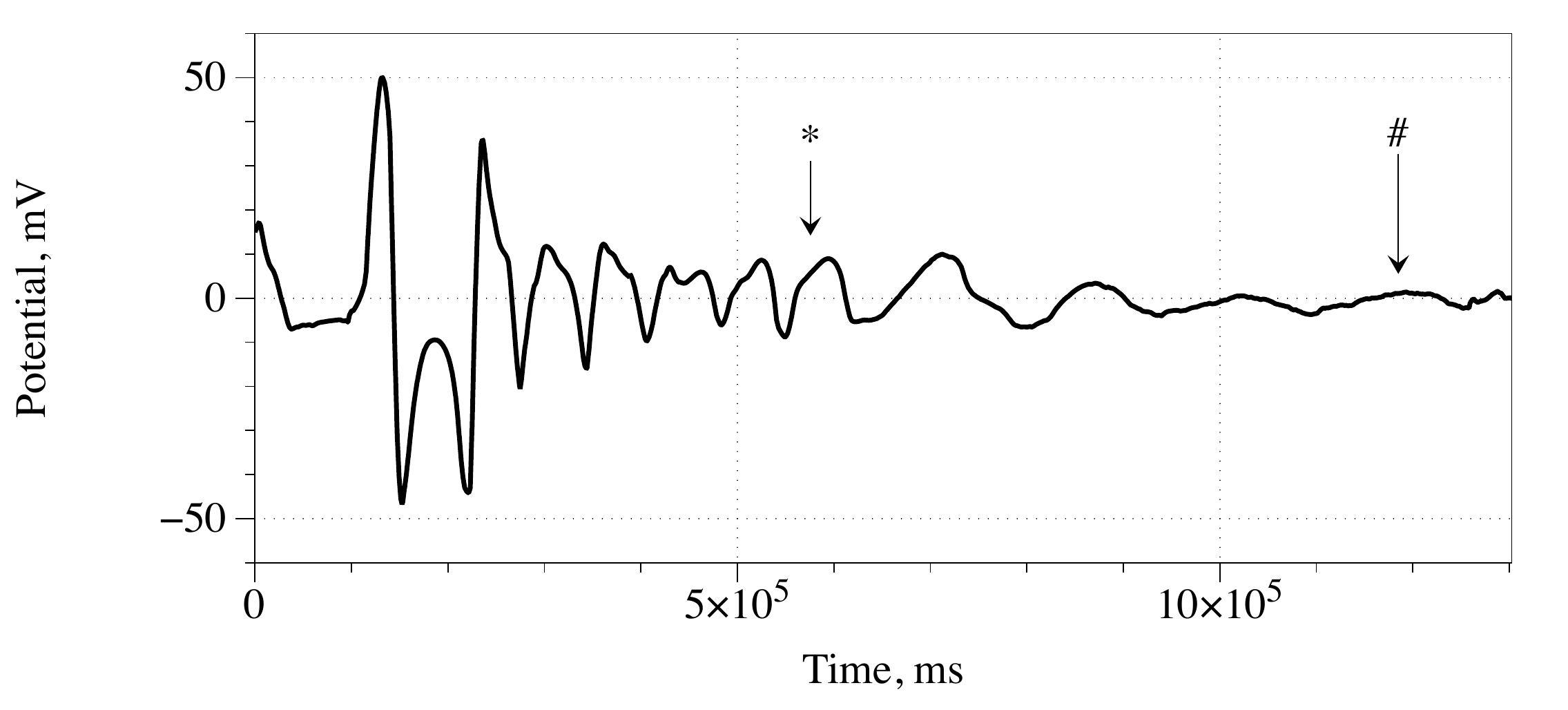}\label{idealpotential}}
      \subfigure[]{\includegraphics[width=0.49\textwidth]{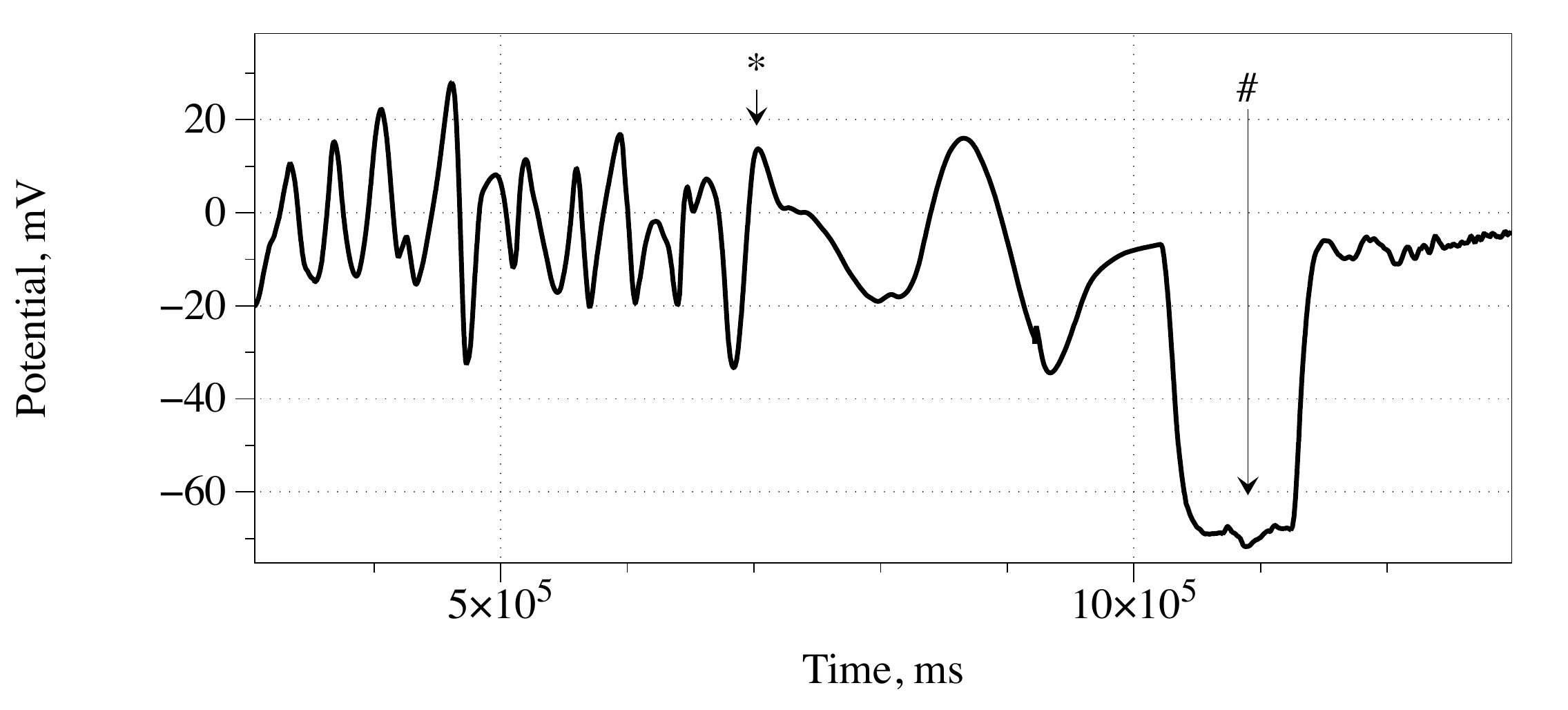}}
    \subfigure[]{\includegraphics[width=0.49\textwidth]{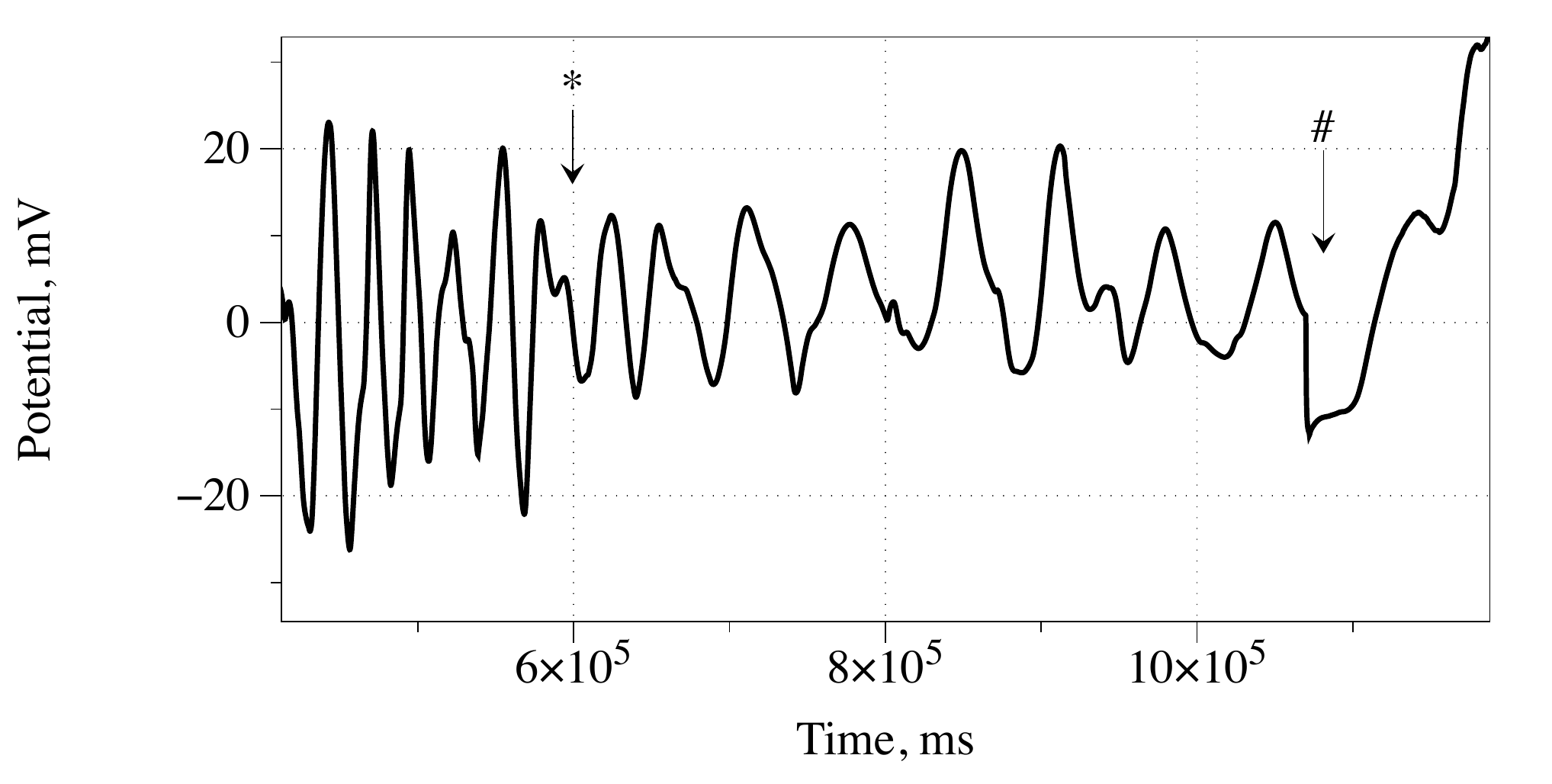}}
    \subfigure[]{\includegraphics[width=0.49\textwidth]{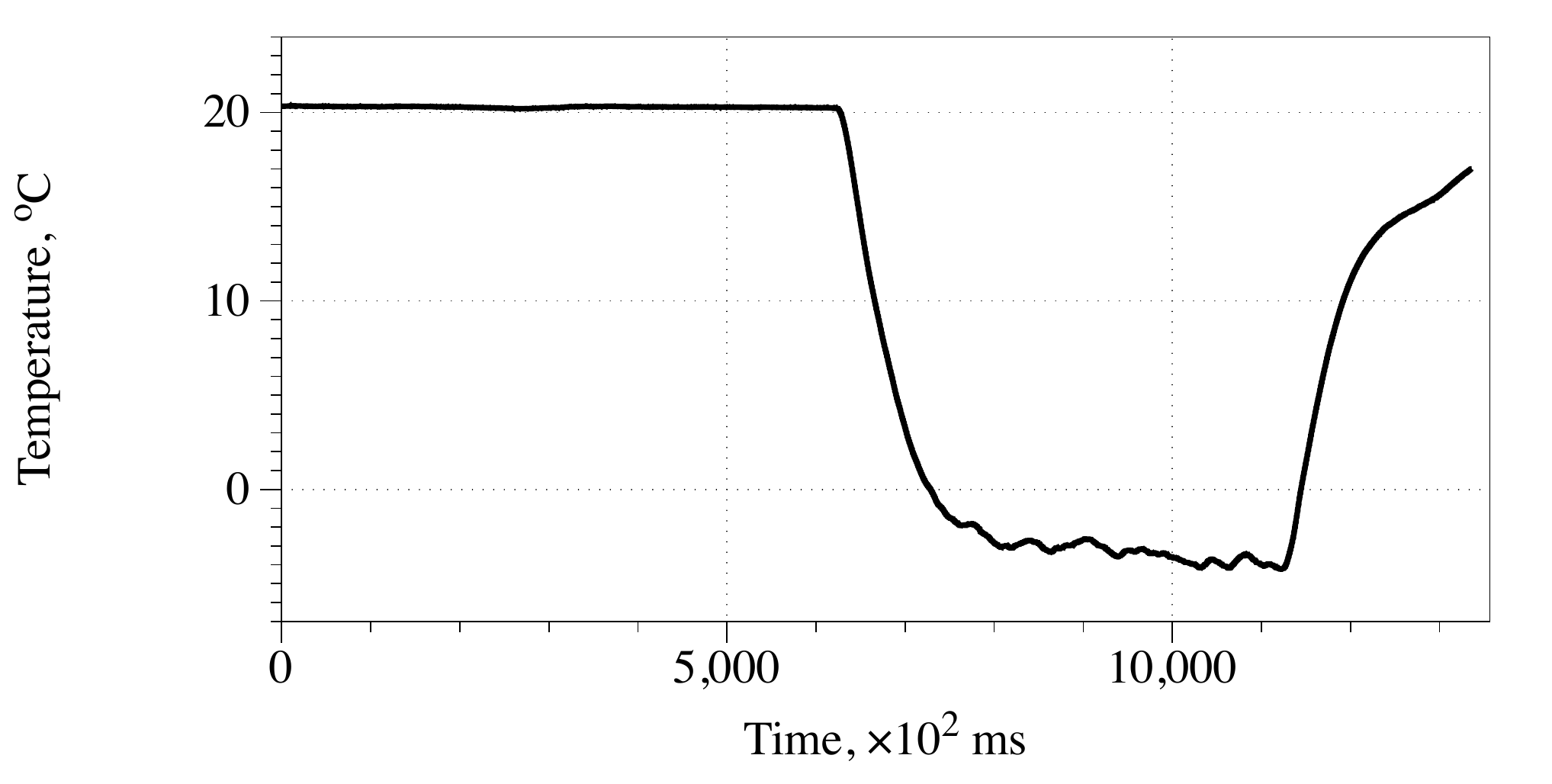}}
    \caption{Dynamics of electrical potential of LM cooled, temperature at the bottom of the Petri dish, down to (a)-4\ce{^o}C, (c)-3\ce{^o}C, (d)~-2\ce{^o}C. Moment when Peltier element is switched on is shown by `*' and off by `\#'. (d)~Temperature log corresponding to experiments (a).}
    \label{fig:burst2}
\end{figure}

\begin{figure}[!tbp]
    \centering
     \subfigure[]{\includegraphics[width=0.49\textwidth]{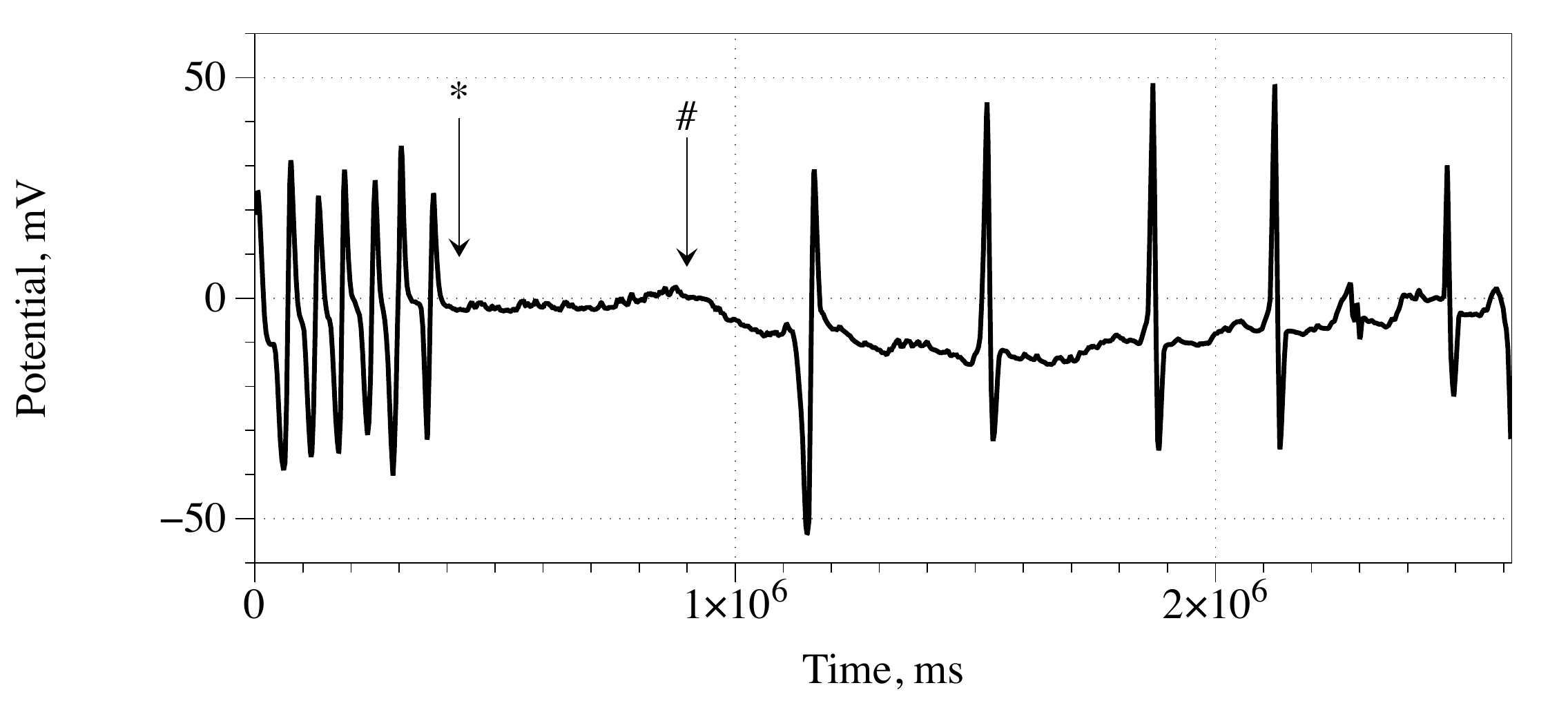}\label{7V_BZLM}}
    \subfigure[]{\includegraphics[width=0.49\textwidth]{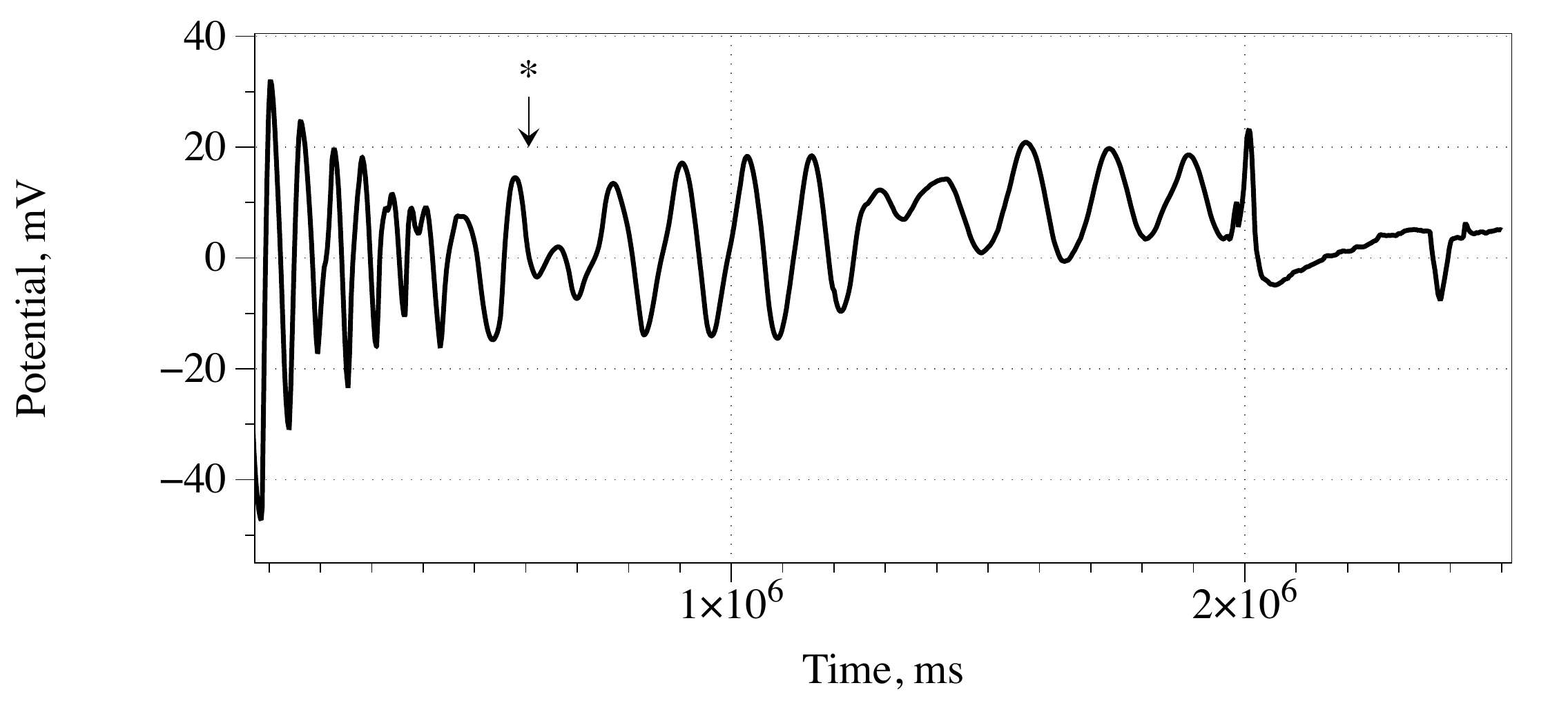}\label{7V_BZLM_2}}
    \subfigure[]{\includegraphics[width=0.49\textwidth]{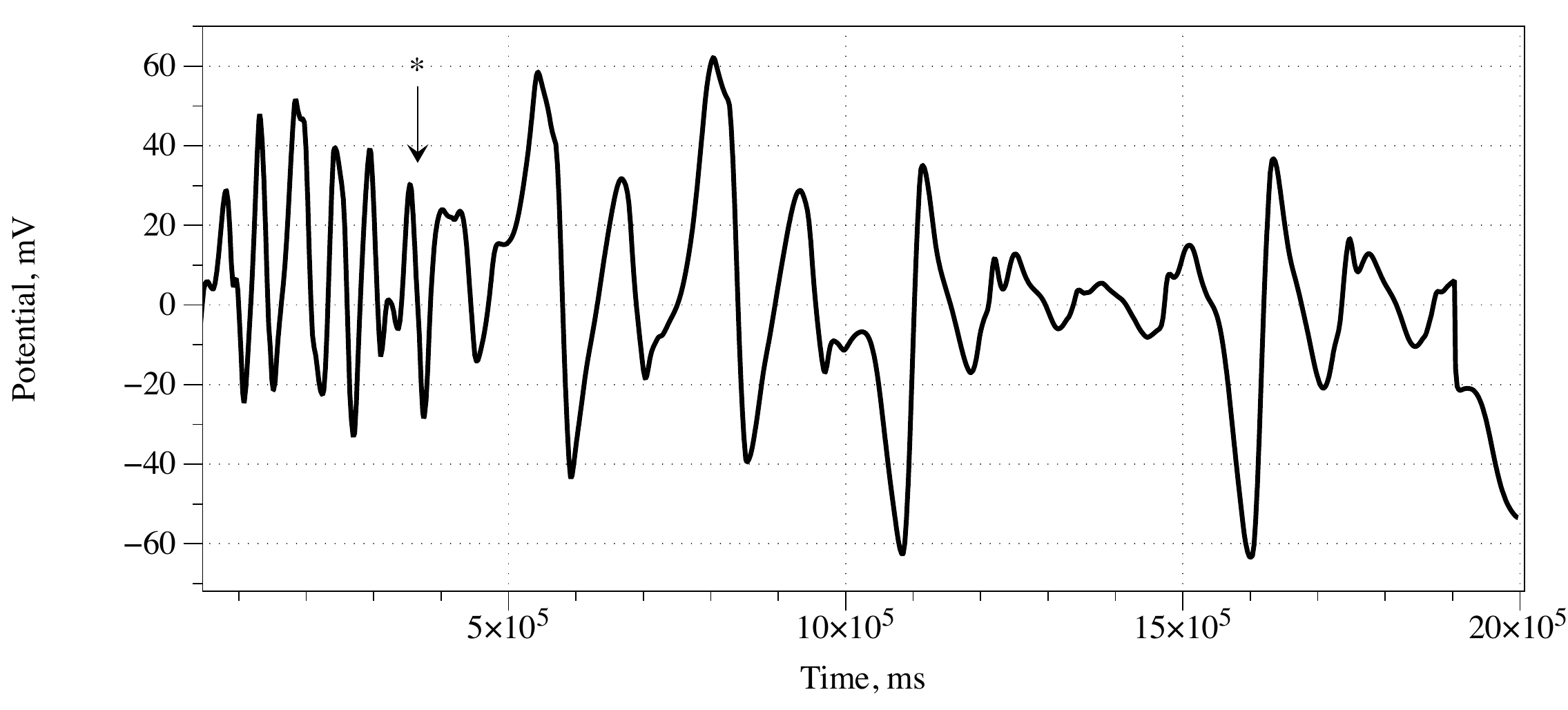}\label{7V_BZ_Marble_061118_5}}
    \subfigure[]{\includegraphics[width=0.49\textwidth]{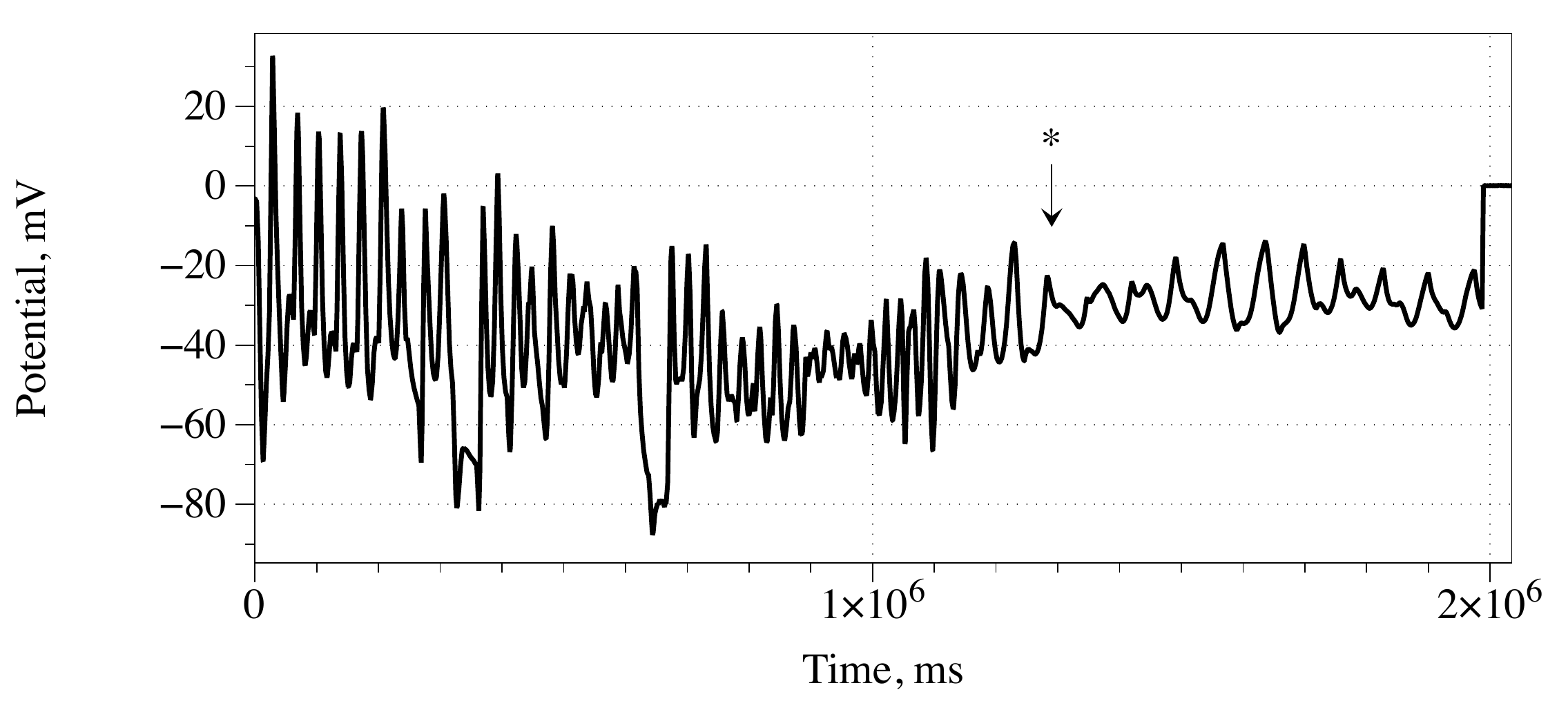}\label{7V_BZ_Freezing_23}}
     \subfigure[]{\includegraphics[width=0.49\textwidth]{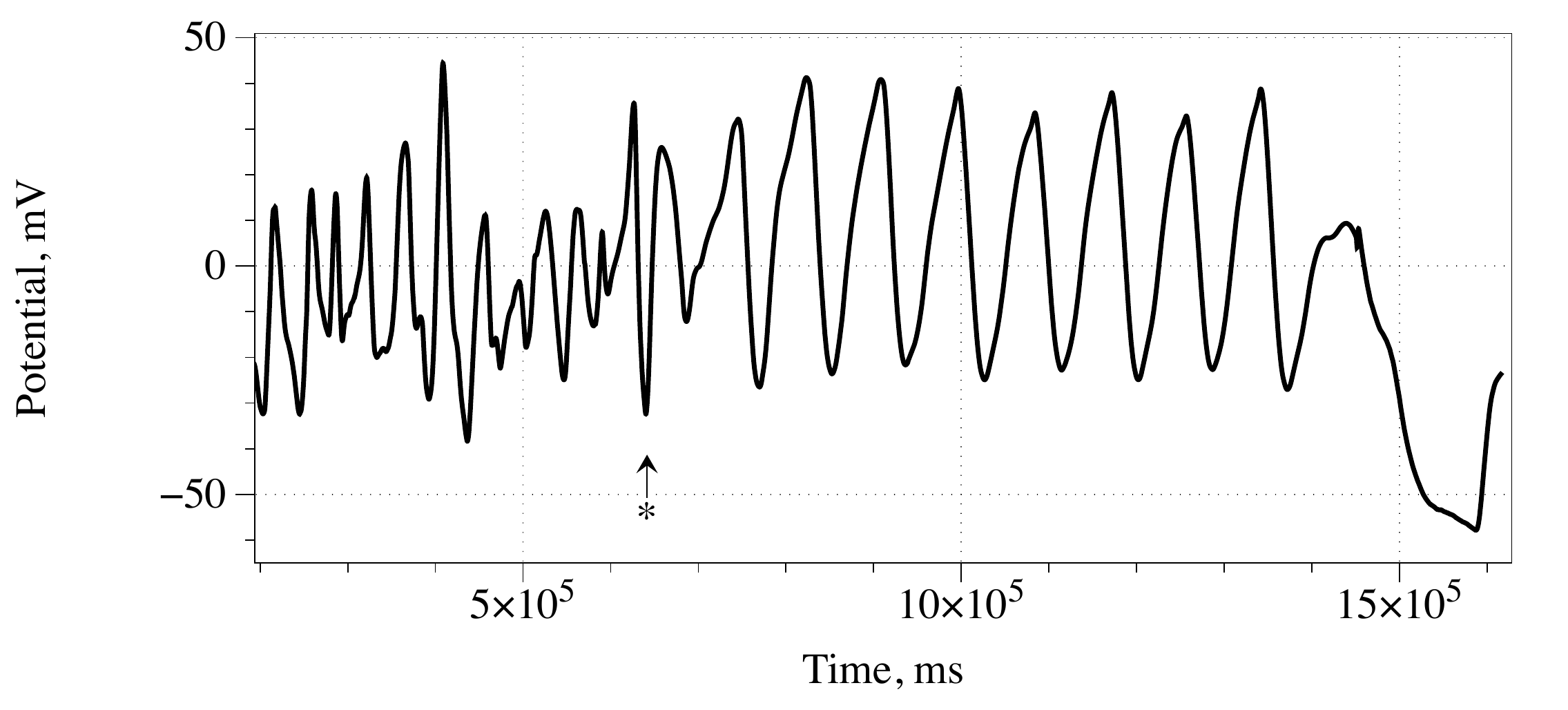}\label{BZ_freezing_7V_7}}
    \subfigure[]{\includegraphics[width=0.49\textwidth]{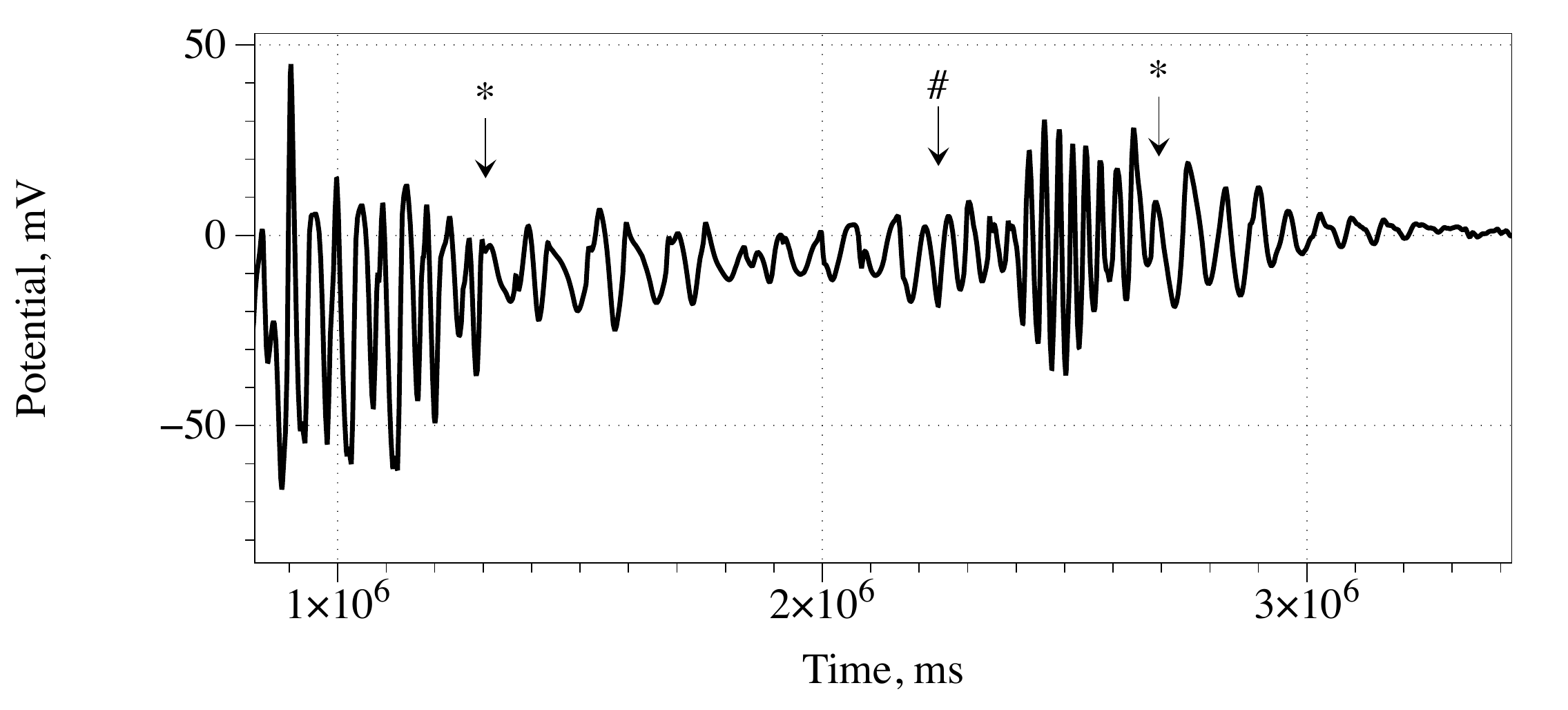}\label{7V_BZ_Freezing_22}} %
    \subfigure[]{\includegraphics[width=0.49\textwidth]{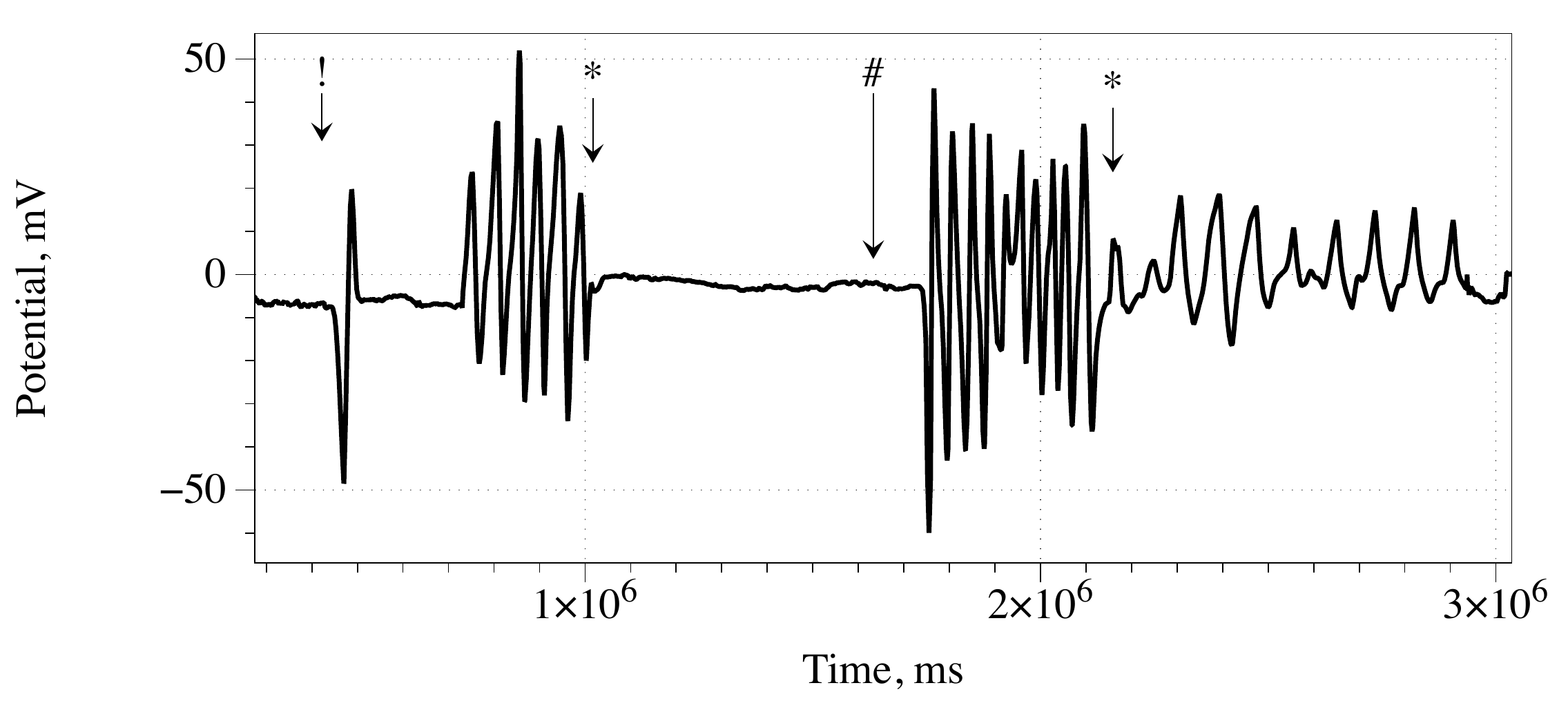}\label{7V_BZLM_3}} %
          \subfigure[]{\includegraphics[width=0.49\textwidth]{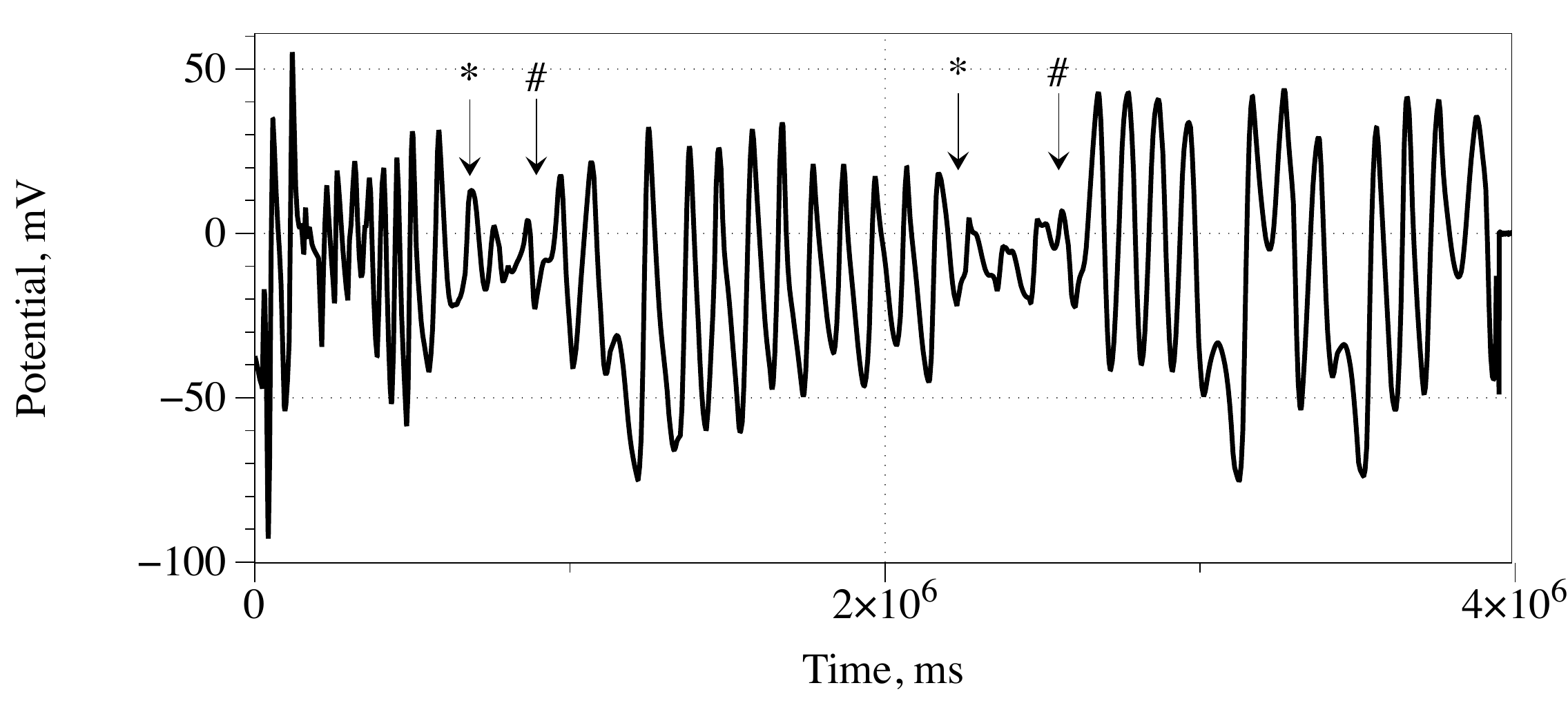}\label{7V_BZ_Marble_061118_1}} %
    \caption{Dynamics of electrical potential of BZ LM subjected to cooling down to -1\ce{^o}C and warming up. (a--e)~LM is cooled down at moment shown by `*' and is never warmed up, (f-g)~LM is cooled down (`*') then warmed up (`\#') then cooled down again (`*'), (h)~a cycle `*\#' of cooling down and warming is repeated twice. Moments when electrodes are inserted in the LM are shown by `!'.}
    \label{fig:7V}
\end{figure}

\begin{table}[!tbp]
    \centering
    \begin{tabular}{c|ccc}
    Plot & $p$, s & $p^*$, s & $\frac{p^*}{p}$ \\ \hline
  Fig.~\ref{7V_BZLM} & 61  & 336 & 5.5 \\
Fig.~\ref{7V_BZLM_2} & 59 & 126 & 2.1\\
Fig.~\ref{7V_BZ_Marble_061118_5} & 56 & 138 & 2.5\\
Fig.~\ref{7V_BZ_Freezing_23} & 22 & 67 &  3\\
Fig.~\ref{BZ_freezing_7V_7} & 39 & 86 & 2.2\\
Fig.~\ref{7V_BZ_Freezing_22} & 47 & 74 & 1.6 \\
Fig.~\ref{7V_BZ_Freezing_22} & 29 & 67 & 2.3 \\
Fig.~\ref{7V_BZLM_3} & 39 & 86 & 2.2
    \end{tabular}
    \caption{Effect of cooling to -1\ce{^o}C on a period of electrical potential oscillations of BZ LM: $p$ is a period of electrical potential oscillation a LM at ambient temperature, $p^*$ is a period of electrical potential oscillations of the cooled LM.}
    \label{tab:frequencies}
\end{table}

Temperatures on the surface of the Petri dish below -2\ce{^o}C usually result in a burst LM. Two examples  are illustrated in Fig.~\ref{fig:burst}. An intact LM (Fig.\ref{intactLM1}) shows  oscillations with average period 26~s (between `A' and `B in Fig.~\ref{plot1}). When cooling is started (`B' in Fig.~\ref{plot1}) oscillations quickly become low frequency low amplitude irregular, average period 49~s. Eventually the LM burst (`C' in  Fig.~\ref{plot1})) and its cargo is relocated away from the electrodes (Fig.~\ref{burstLM1}). In scenario  shown in Fig.~\ref{intactLM2}--\ref{plot2}  LM undergoes two instances of freezing. First time, marked `B' in Fig.~\ref{plot2} the LM (Fig.~\ref{intactLM2}) survives being cooled down with just slight change in shape (Fig.~\ref{FreezingLM2}). Period of oscillations increases from 28~s in intact LM to 162~s in cooled down LM (period between `B' and 'C' in Fig.~\ref{plot2}). After Peltier is switched off (moment `C' in Fig.~\ref{plot2}) the LM resumes high frequency oscillations, frequency 42~s, but with lower amplitude. The LM does not survive second round of freezing (`D' in Fig.~\ref{plot2}) and bursts, whilst still wetting the electrodes (Fig.~\ref{BurstLM2}). More examples of electrical potential dynamics for temperatures causing LM bursting are shown in Fig.~\ref{fig:burst2}. The temperature of -2\ce{^o}C is critical, in that over 70\% of LMs burst and did not survive second round of freezing. Therefore, in further experiments the LMs were cooled down to -1\ce{^o}C.

Patterns of oscillations of LM cooled down to -1\ce{^o}C show a high degree of polymorphism (Fig.~\ref{fig:7V}) in amplitudes. Changes in frequencies are as follows are analysed in Tab.~\ref{tab:frequencies}. If we ignore the first example (Fig.~\ref{7V_BZLM}) then we have average $p=44.4$ ($\sigma(p)=12.5$), average $p^*=92$ ($\sigma(p^*)=28.6$) average $\frac{p^*}{p}=2.1$ ($\sigma(\frac{p^*}{p})=0.5$).

In the experiments shown in Figs.~\ref{7V_BZLM}, \ref{7V_BZLM_2}, \ref{7V_BZ_Marble_061118_5}, \ref{7V_BZ_Freezing_23}, and \ref{BZ_freezing_7V_7}, LMs were kept cooled until the end of the experiments. In the experiment shown in Fig.~\ref{7V_BZ_Freezing_22}, cooling was started after 1310 seconds of the experiment, the Peltier was switched off after 2254 seconds, and cooling was repeated at 2690 seconds. Intact LM oscillated with average period 47~s at first phase of the experiment. Cooled LM oscillated with period 74~s. The period became 29~s after the warming. Second cooling increased the period to 67~s. Thus, we have an increase of 1.6 times during first cooling cycle and by 2.3 times during the second cooling cycle. In the experiment illustrated in Fig~\ref{7V_BZLM_3}, oscillations were arrested by cooling yet restarted when the LM was warmed. Period of oscillations before cooling was 47~s, and after oscillations restarted after cooling was 39~s. Repeated cooling did not arrest oscillations yet increased the oscillation period 2.2 times to 86~s. In the experiment shown in Fig.~\ref{7V_BZ_Marble_061118_1}, we cooled a LM for short periods of time (199~s and 288~s) and did not observe any substantial changes in periods of oscillation, after the first freezing cycle. The average periods were changing as follows 46~s $\rightarrow$ 92~s $\rightarrow$ 98~s $\rightarrow$ 98~s $\rightarrow$ 98~s.  

To summarise, average period of oscillations of a BZ LM doubles from 44~s to 92~s when the LM is cooled down to -1\ce{^o}C. The frequency of oscillations is restored after cooling is stopped. The amplitude of oscillations may increase or decrease as a result of cooling. Sometimes the oscillations can be completely arrested yet resume after warming.

\section{Discussion}
\label{discussion}

Why are oscillations of electrical potential observed? The oxidation of malonic acid by bromate ions in acidified solution is catalysed by ferroin ions. Ferroin ions \ce{[Fe(o-phen)_3]^{2+}} are oxidised  to their ferric derivatives \ce{[Fe(o-phen)_3]^{3+}}. The ratios of ferroin to ferric ions and bromid ions oscillate in time. This is is reflected in the oscillations of the electrical potential recorded from the LM. If the BZ solution in a LM was mixed, then global oscillations would occur, resulting in the potential at both electrodes being the same and therefore no electrical oscillations could be observed. However the solution is not mixed, therefore waves of oxidation emerge spontaneously, or are induced when the LM is pierced by electrodes, or induced by piercing with a silver wire.  Therefore the ratio of ferroin to ferric ions (and bromide ions) are changing only at the wave front. Thus when the wave front passes the electrodes the electrical potential difference is observed.

\begin{figure}[!tbp]
    \centering
\subfigure[]{\includegraphics[width=0.24\textwidth]{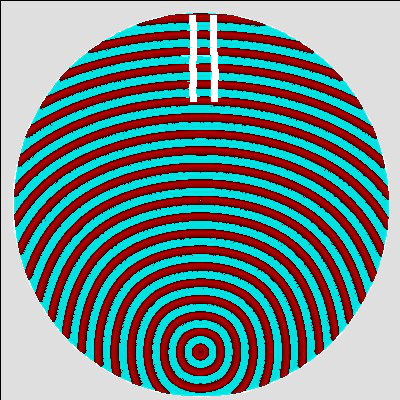}\label{Swave}}
\subfigure[]{\includegraphics[width=0.55\textwidth]{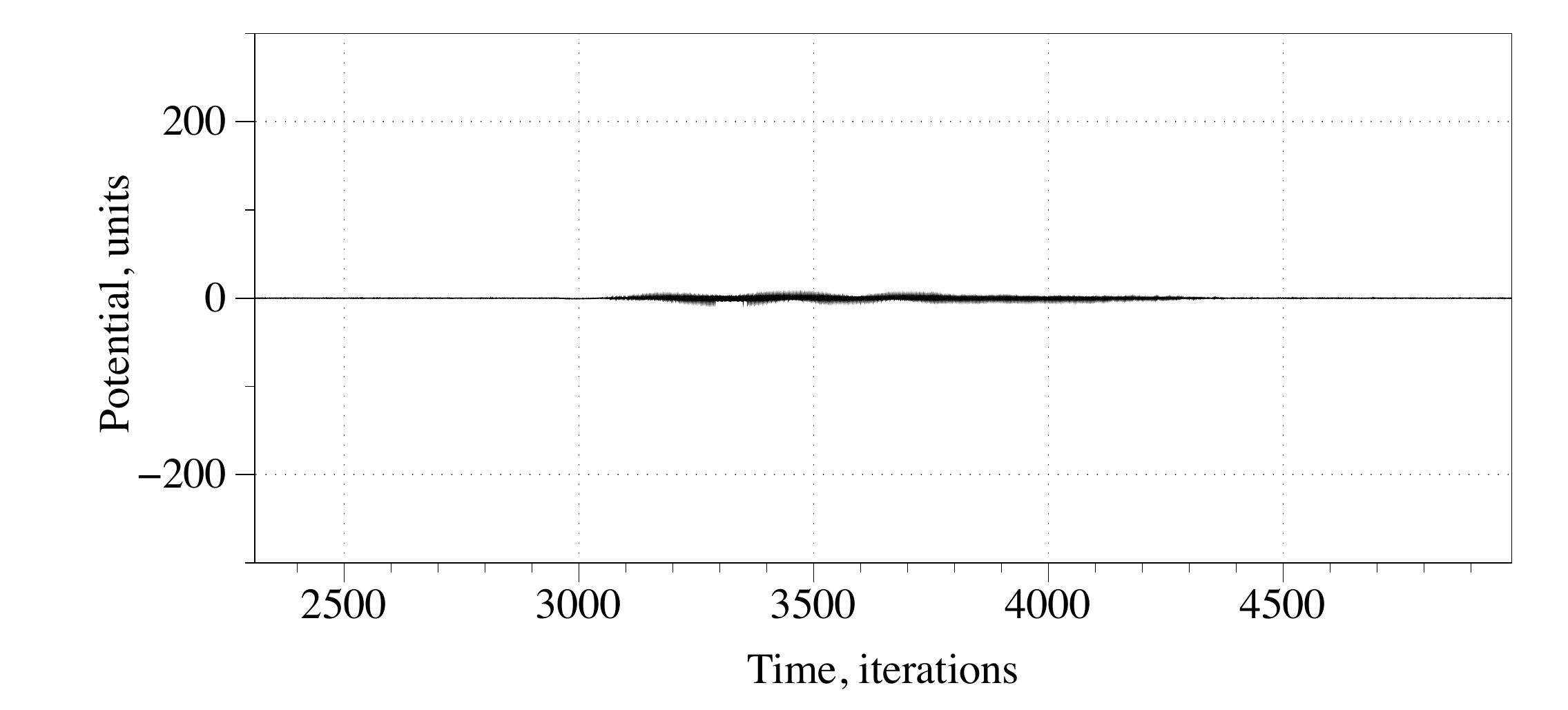}\label{Swaveplot}}
\subfigure[]{\includegraphics[width=0.24\textwidth]{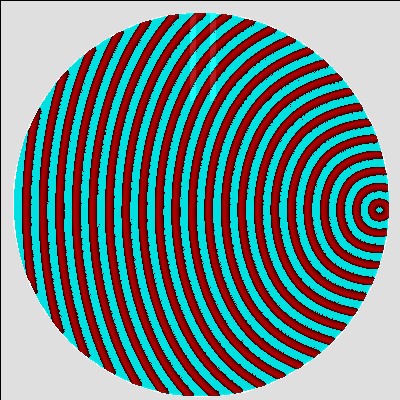}\label{Ewave}}
\subfigure[]{\includegraphics[width=0.55\textwidth]{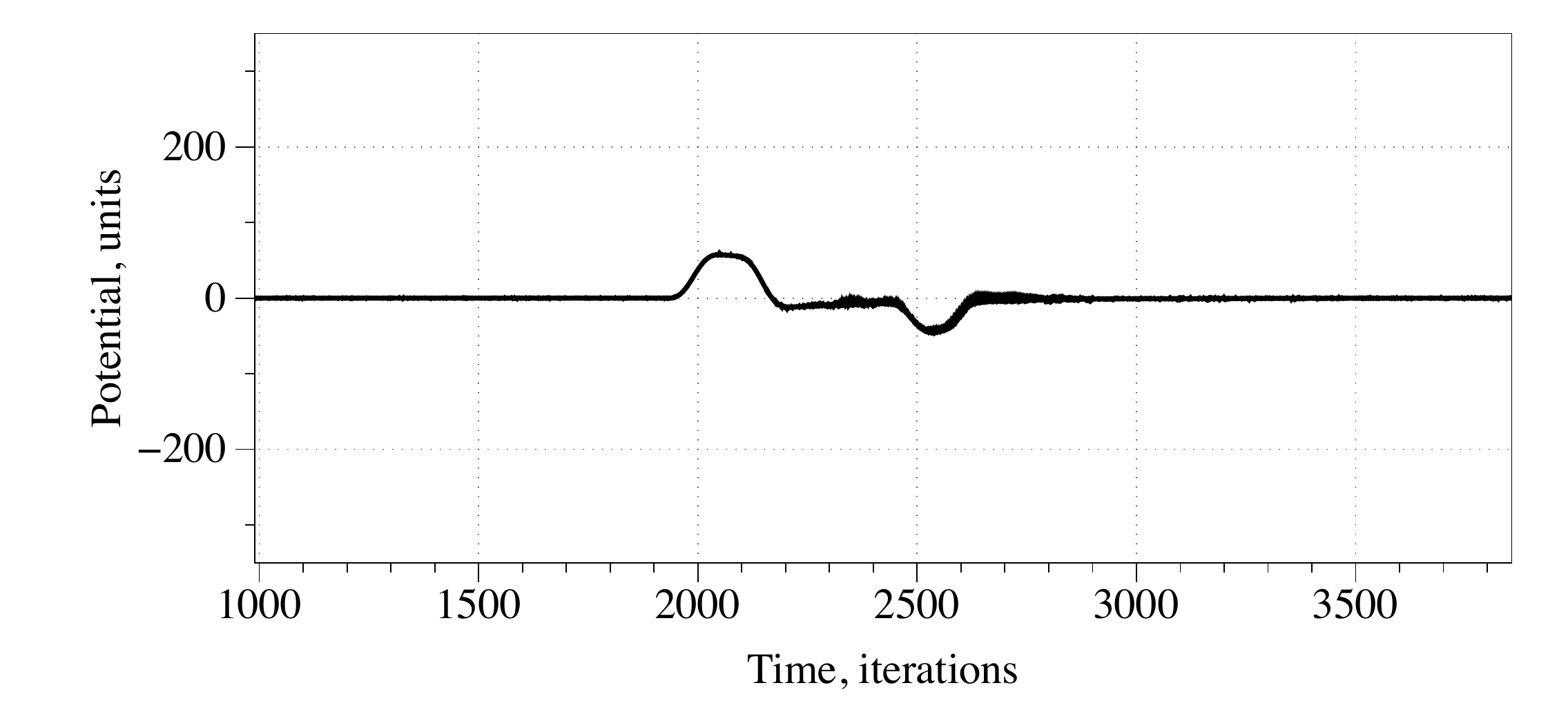}\label{Ewaveplot}}
\subfigure[]{\includegraphics[width=0.24\textwidth]{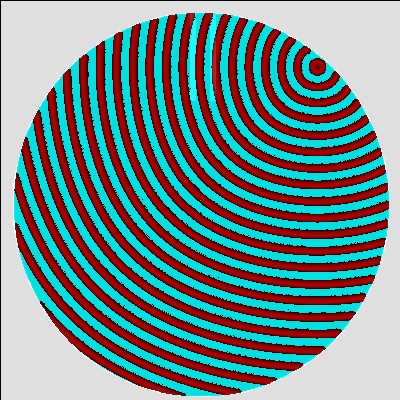}\label{NEwave}}
\subfigure[]{\includegraphics[width=0.55\textwidth]{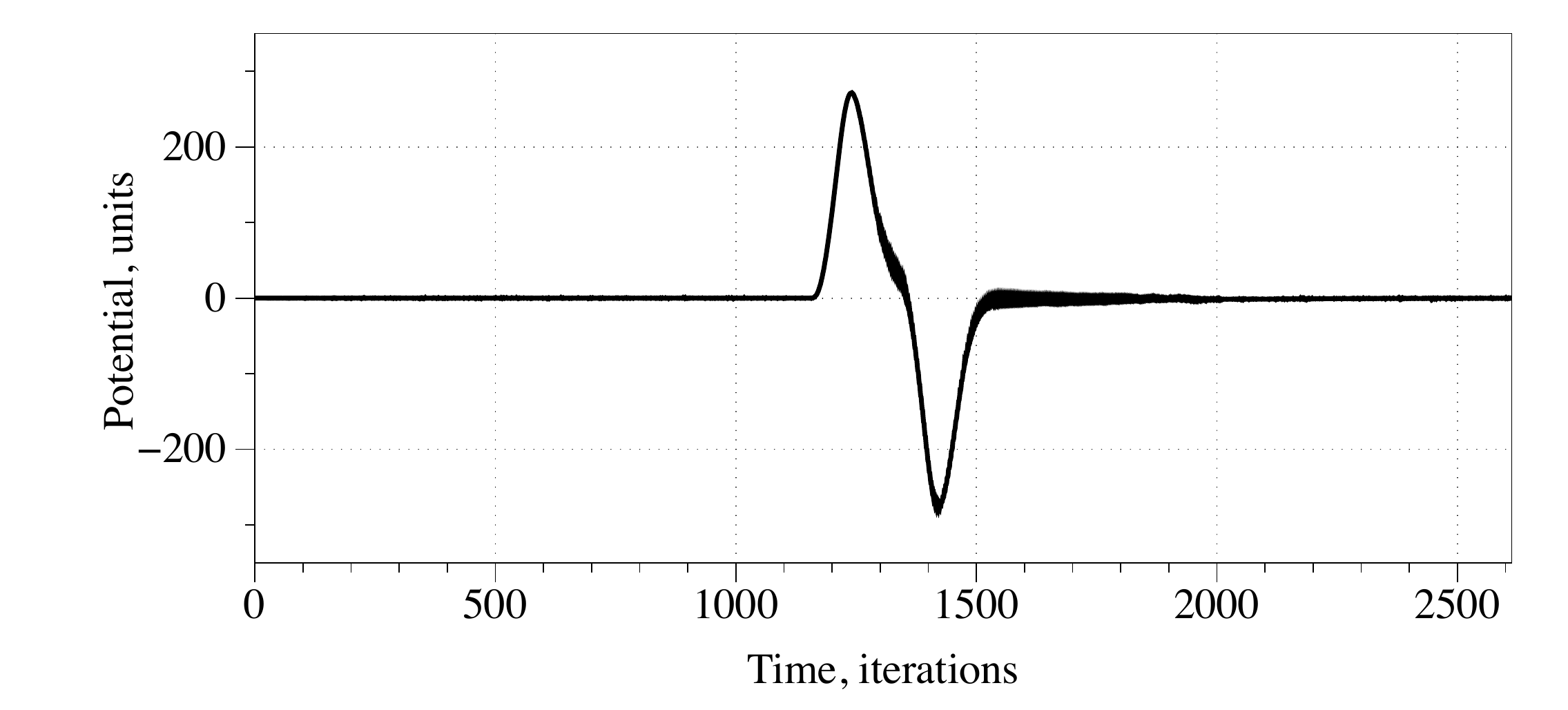}\label{NEwaveplot}}
    \caption{
    Time-lapse snapshots (ace) and corresponding potential recorded at the electrodes (bdf) of a single-wave initiated at southern edge of the droplet (ab), Eastern edge of the droplet (cd), north-eastern edge of the droplet (ef).
    The time-lapse snapshots  were recorded at every 150\textsuperscript{th} time step.  We display sites with $u >0.04$. Domains corresponding to the electrodes are shown by white rectangles in (a).
    }
    \label{fig:differentspikes}
\end{figure}

Why patterns of oscillations are not always regular? This is because several oxidation waves, and even several generators/sources of oxidation waves can co-exist in a single LM. These waves can superimpose with each other, collide and annihilate in the result of the collisions, or produce localised wave-fragments. This rich dynamic of wave-fronts is reflected in, sometimes, irregular patterns of oscillation. Let us illustrate further discussions with a two-variable Oregonator equations~\cite{field1974oscillations,beato2003pulse}:

\begin{eqnarray}
  \frac{\partial u}{\partial t} & = & \frac{1}{\epsilon} (u - u^2 - (f v + \phi)\frac{u-q}{u+q}) + D_u \nabla^2 u \nonumber \\
  \frac{\partial v}{\partial t} & = & u - v.
\label{equ:oregonator}
\end{eqnarray}

\begin{figure}[!htbp]
    \centering
     \subfigure[$\phi=0.01$, $t=10^4$]{\includegraphics[width=0.24\textwidth]{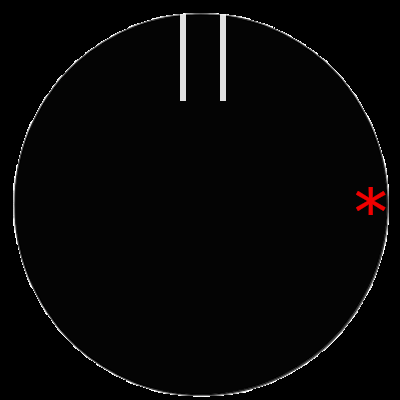}\label{electrodes}}
    \subfigure[$\phi=0.01$, $t=10^4$]{\includegraphics[width=0.24\textwidth]{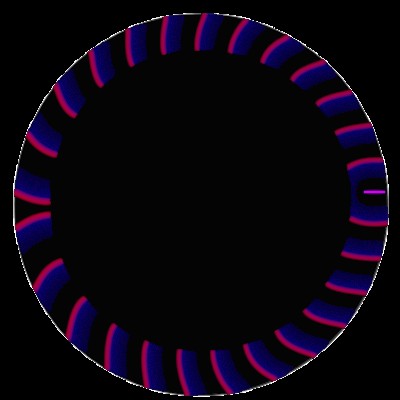}\label{snapshot_fi001}}
    \subfigure[$\phi=0.03$,$t=10^4$]{\includegraphics[width=0.24\textwidth]{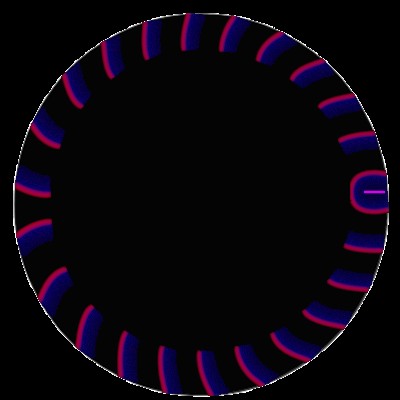}\label{snapshot_fi003}}
\subfigure[$\phi=0.05$,$t=10^4$]{\includegraphics[width=0.24\textwidth]{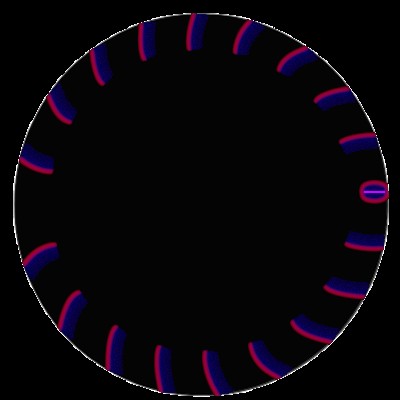}\label{snapshot_fi005}}
\subfigure[$\phi=0.07$,$t=3\cdot 10^4$]{\includegraphics[width=0.24\textwidth]{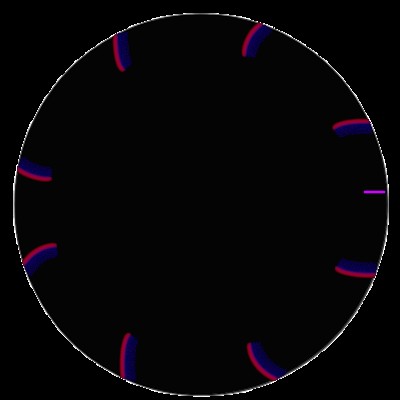}\label{snapshot_fi007}}
\subfigure[$\phi=0.01$]{\includegraphics[width=0.49\textwidth]{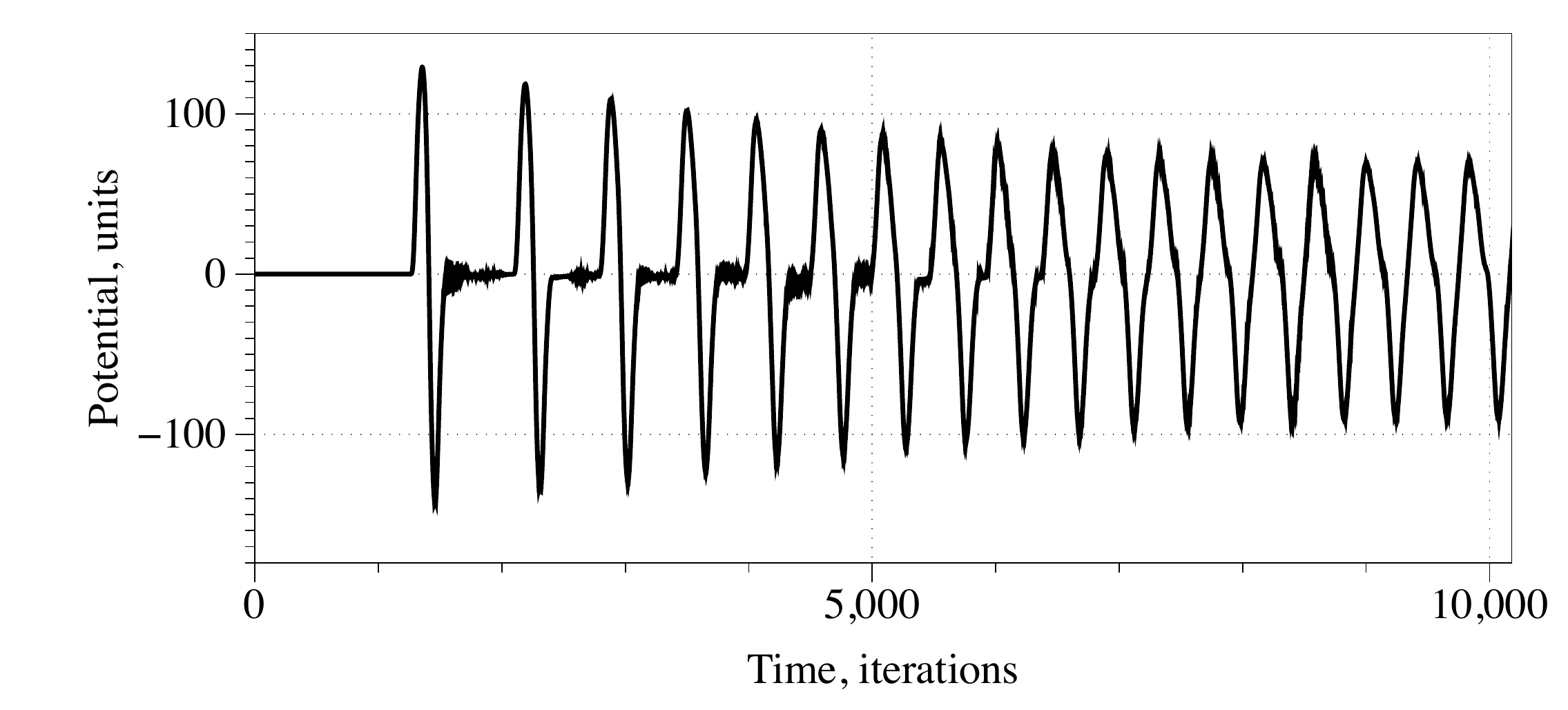}\label{oscillations_fi001}}
\subfigure[$\phi=0.03$]{\includegraphics[width=0.49\textwidth]{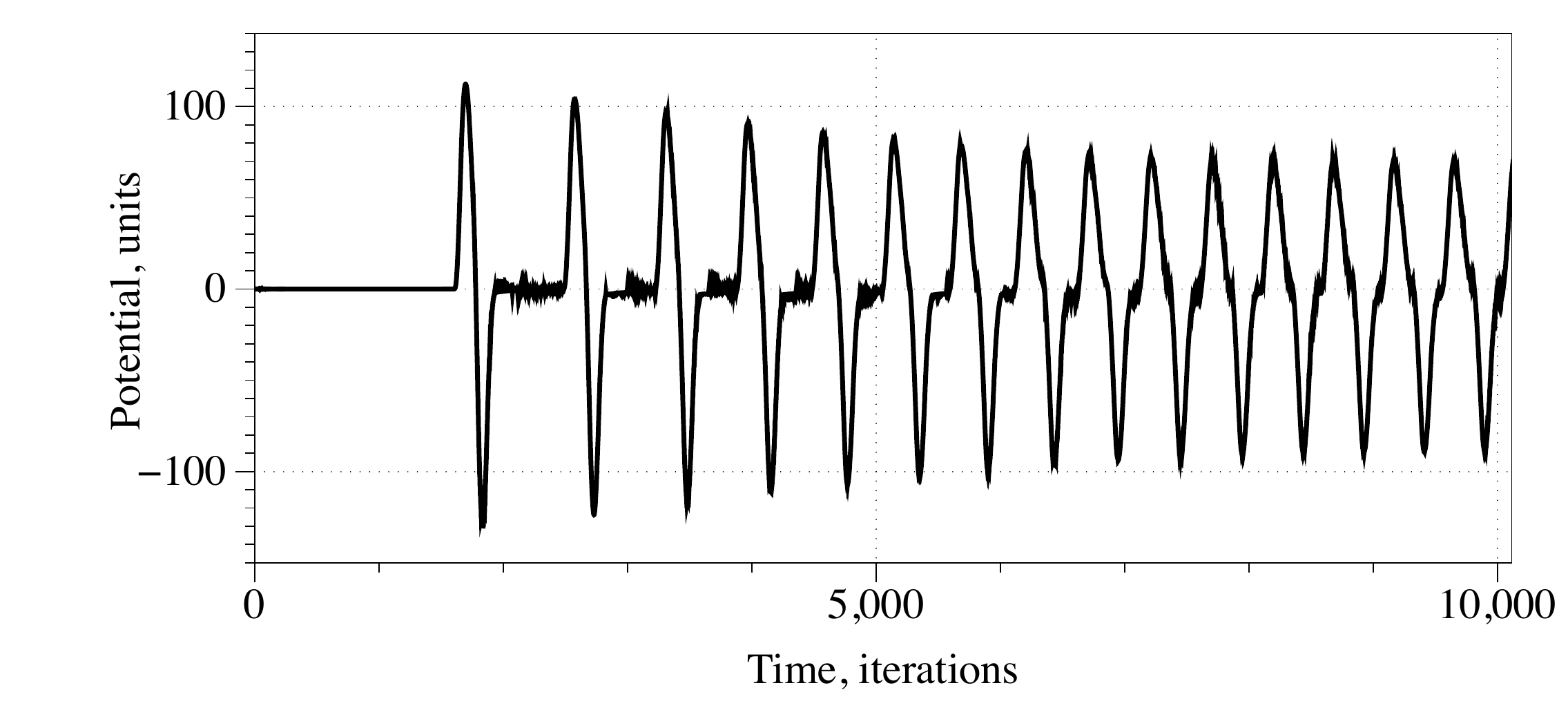}\label{oscillations_fi003}}
\subfigure[$\phi=0.05$]{\includegraphics[width=0.49\textwidth]{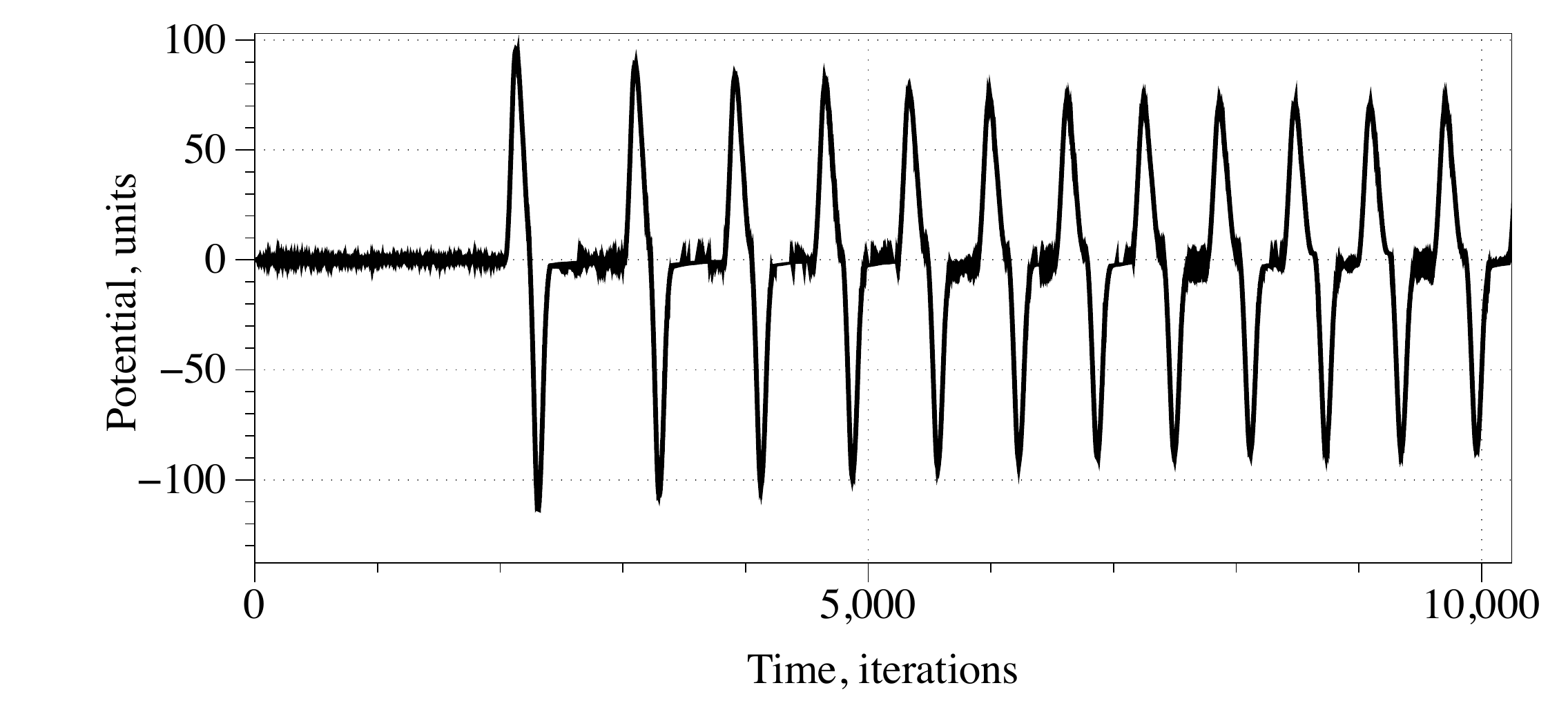}\label{oscillations_fi005}}
\subfigure[$\phi=0.07$]{\includegraphics[width=0.49\textwidth]{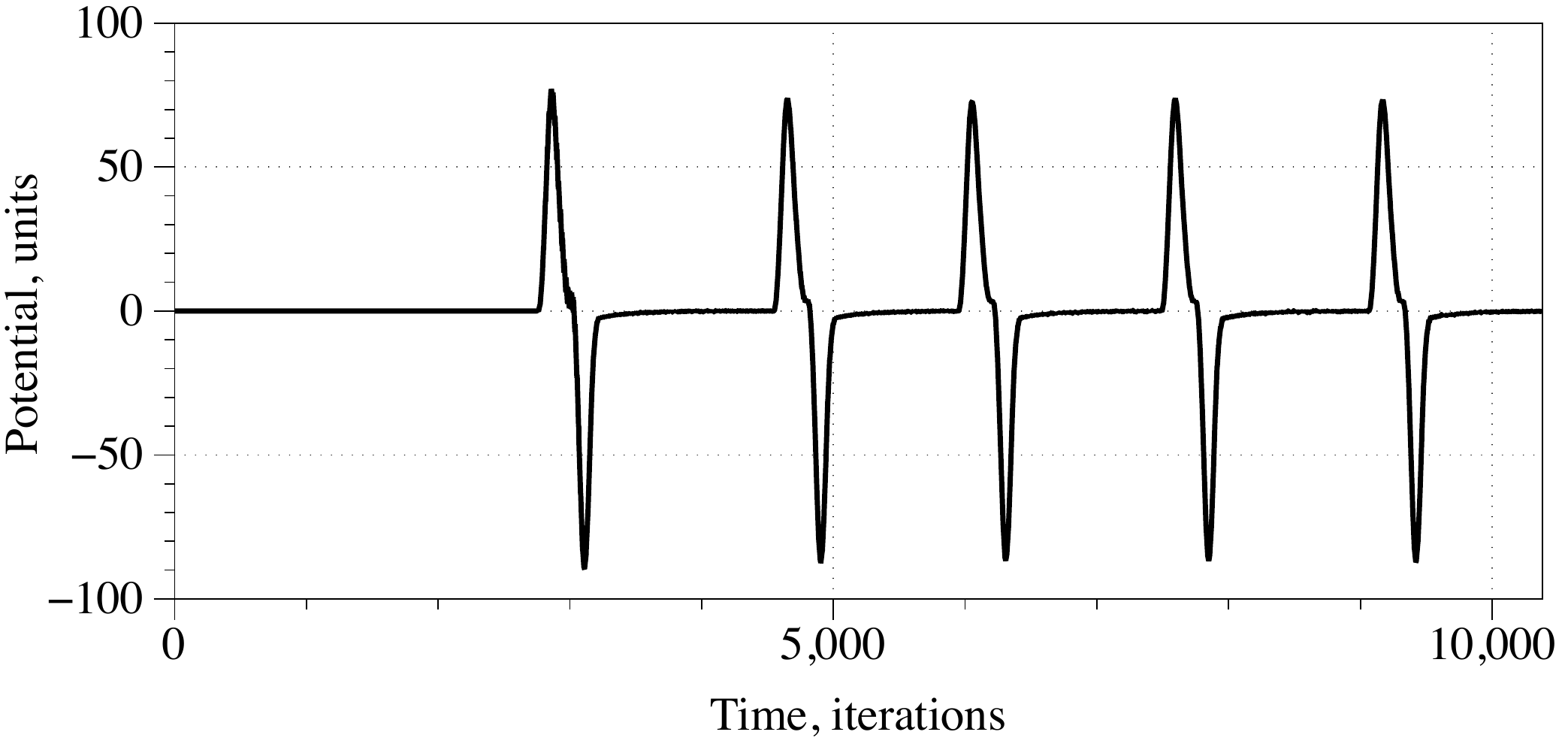}\label{oscillations_fi007}}
    \subfigure[]{\includegraphics[width=0.4\textwidth]{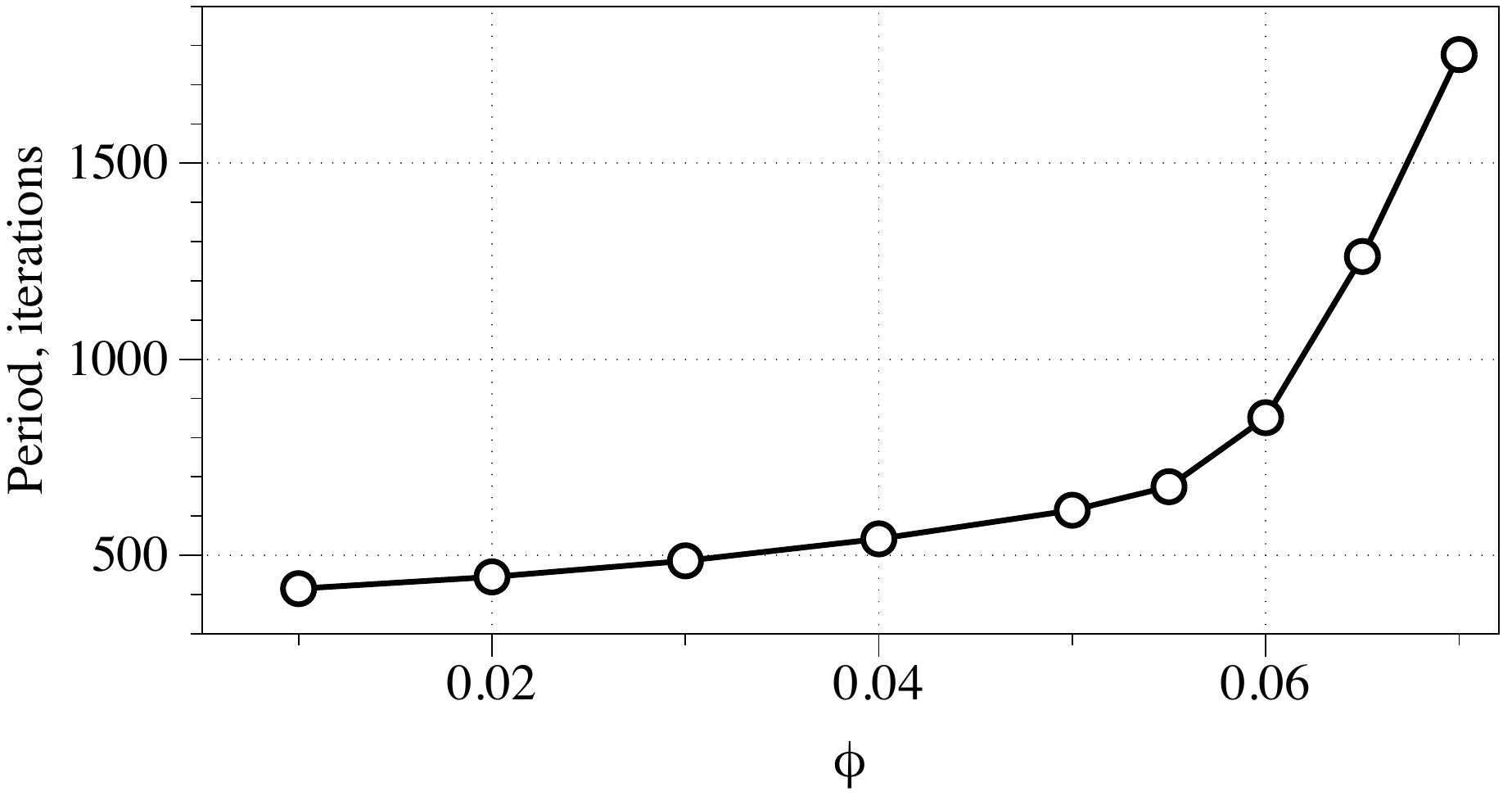}\label{Period_vs_Fi}}
    \caption{Modelling wave-front propagation and frequency of wave-generation for various values of excitability $\phi$.
    (a)~Position of electrodes, auto-excitation loci $\mathcal{L}$ is shown by star. 
    (b--e)~Snapshots of the medium showing trains of the wave-fronts, values of $\phi$ are shown in sub-captions. 
    (f--i)~Potential difference recorded on the electrodes, values of $\phi$ are shown in sub-captions. 
    (j)~Period of the potential oscillations versus excitability $\phi$.
    }
    \label{fig:frequency_vs_phi}
\end{figure}

 The variables $u$ and $v$ represent local concentrations of an activator, or an excitatory component of BZ system, and an inhibitor, or a refractory component. Parameter $\epsilon$ sets up a ratio of the time scale of variables $u$ and $v$, $q$ is a scaling parameter depending on rates of activation/propagation and inhibition, and $f$ is a stoichiometric coefficient. We integrated the system using Euler method with five-node Laplace operator, time step $\Delta t=0.001$ and grid point spacing $\Delta x = 0.25$, $\epsilon=0.02$, $f=1.4$, $q=0.002$. We varied value of $\phi$ from the interval $\Phi=[0.05,0.08]$, where constant $\phi$ is a rate of inhibitor production. $\phi$ represents the rate of inhibitor; this rate can be dependent on light, temperature, or presence of other chemical species. The parameter $\phi$ characterises excitability of the simulated medium, i.e. the larger $\phi$ the less excitable the medium is. We represent BZ LM as a disc with a radius of 185 nodes. We represent electrodes as rectangular domains of the discs, see Fig.~\ref{Swave}  and Fig.~\ref{electrodes}, $\mathcal{E}_1$ and $\mathcal{E}_2$. We calculate the potential difference at each iteration $t$ as 
 $\sum_{x \in \mathcal{E}_2} u^t_x - \sum_{x \in \mathcal{E}_1} u^t_x$. 
 
 Orientation of the wave-front passing the electrodes determines exact shape of the impulse recorded (Fig.~\ref{fig:differentspikes}). Assume a droplet is excitable everywhere.  If a wave-front is perpendicular to the electrodes, e.g. a wave is generated at the southern edge of the droplet (Fig.~\ref{Swave}), the potential difference between electrodes at any moment of time will be near zero, a part of some noise (Fig.~\ref{Swaveplot}). A wave originated at the eastern edge of a droplet enters electrodes at an obtuse angle (Fig.~\ref{Ewave}). This is reflected in two spikes --- one is positive potential and another is negative potential (Fig.~\ref{Swaveplot}), there is a substantial distance between the spikes.  If the wave-front propagates nearly parallel to the electrodes, e.g. when a wave is generated at north-east edge of the droplet (Fig.~\ref{NEwave}), the action-like potential is recorded (Fig.~\ref{NEwaveplot}), which shape imitates distinctive depolarisation, repolarisation and hyperpolarisation phases of a biological action potential. In experiments we always observed oscillation. The shape of the impulses was nearly the same --- subject to deviations --- in all experiments. This implies that the wave-front travels not in the volume of BZ LM but along the surface of the LM. Thus the wave-front passes electrodes being nearly parallel to them.  %Not sure on this last bit. (Tom)

\begin{figure}[!htbp]
    \centering
    \subfigure[90~s]{\includegraphics[width=0.3\textwidth]{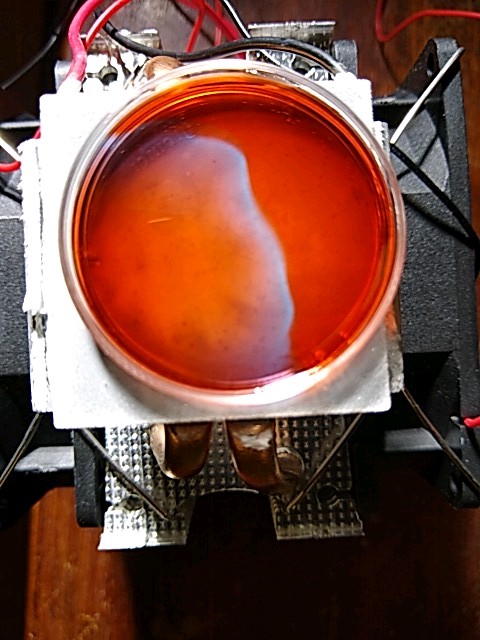}\label{90s}}
    \subfigure[126~s]{\includegraphics[width=0.3\textwidth]{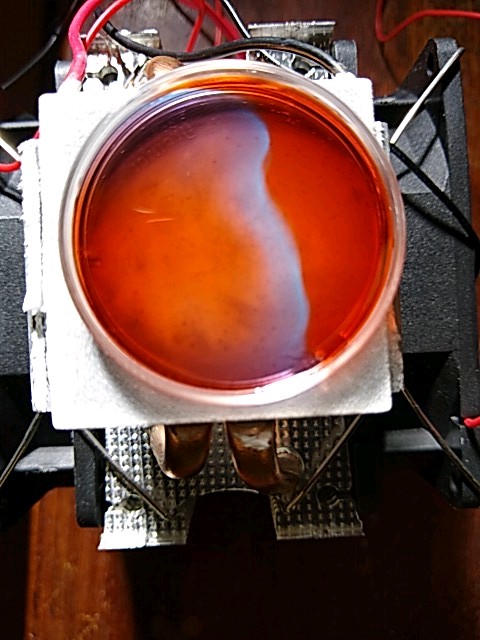}\label{126s}}
    \subfigure[150~s]{\includegraphics[width=0.3\textwidth]{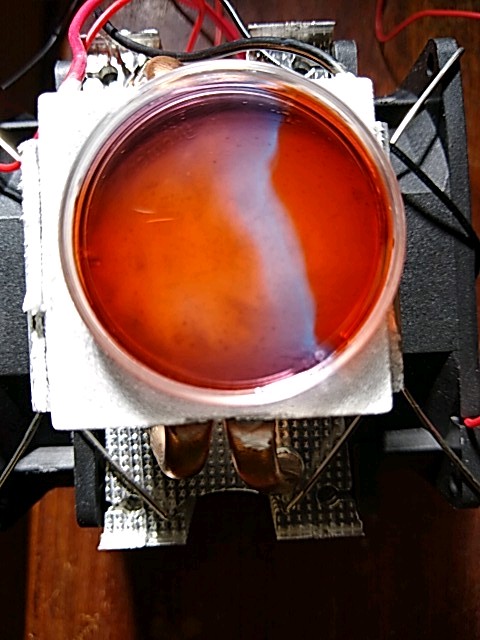}\label{150s}}
    \subfigure[174~s]{\includegraphics[width=0.3\textwidth]{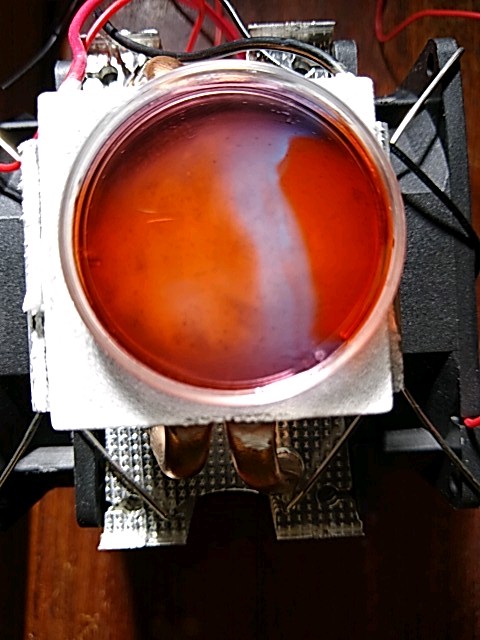}\label{174s}}
    \subfigure[198~s]{\includegraphics[width=0.3\textwidth]{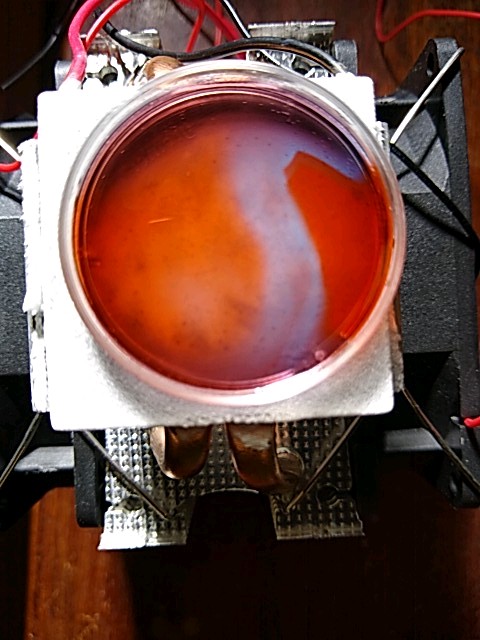}\label{198s}}
\subfigure[246~s]{\includegraphics[width=0.3\textwidth]{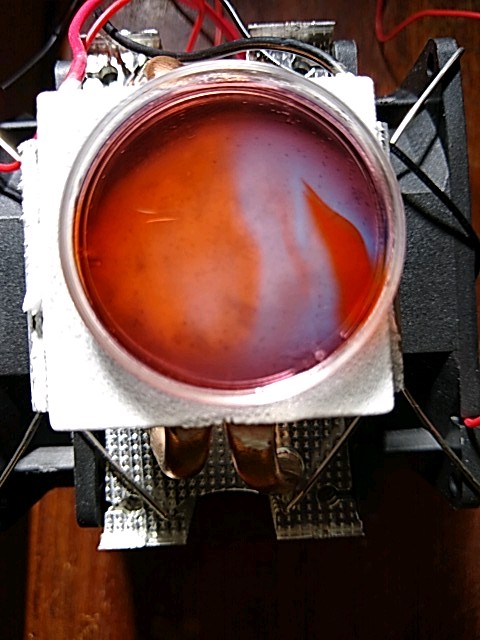}\label{246s}}
    \caption{Time lapse photos of the propagation of an oxidation wave-front in a thin layer of BZ medium on the freezing Peltier element. Time from the start of recording is shown in captions.}
    \label{fig:freezingfront}
\end{figure}

Why does the frequency of oscillations decrease on cooling? Temperature changes the rate of the reaction which consumes the inhibitor of the auto-catalytic \ce{Br^-}~\cite{blandamer1975investigation} species. When the temperature decreases the rate of consumption of \ce{Br^-} also decreases, which increases the time necessary for the reaction to enter its auto-catalytic step. The enlargement of the refractory tail reduces the number of wave-fronts that can be fitted in a limited space. Thus less waves pass electrodes in a given period of time. This is reflected in a reduced frequency of oscillations. The mechanism is illustrated in experiments with a thin-layer BZ medium shown in Fig.~\ref{fig:freezingfront} and simulation with Oregonator model in Fig.~\ref{fig:frequency_vs_phi}.  A 35~mm Petri dish was placed on the freezing setup (Fig.~\ref{fig:setup}) and the element was chilled to -7\textsuperscript{o}C. The BZ medium did not freeze but its temperature dropped to near 0\textsuperscript{o}C. The cooling was reflected in the enlarged tail of the excitation wave front, it doubled in width from 2.5~mm (Fig.~\ref{90s}) to 4.7~mm (Fig.~\ref{198s}) in just over 3~min.  In modelling the BZ medium (Fig.~\ref{fig:frequency_vs_phi}) we position electrodes in North of the droplet and assume that a self-excitation loci near the edge at the East of the droplet (Fig.~\ref{electrodes}) and assume that waves propagate only near the surface (i.e. only part of of 370 nodes wide disc with $r>150$ is excitable).  The excitable loci $\mathcal{L}$ have values $u_x=1$, $x \in \mathcal{L}$, at every iteration of the numerical integration however waves are generated only with some intervals. Distance between wave-fronts increases with decrease of excitability, increase of $\phi$ from 0.01 (Fig.~\ref{snapshot_fi001}) to 0.07 (Fig.~\ref{snapshot_fi007}). This is reflected in decreasing of oscillation frequency of the potential difference recorded at the electrodes (Fig.\ref{oscillations_fi001}--\ref{oscillations_fi007}). The shapes of impulses in Fig.~\ref{oscillations_fi007} strikingly resemble shapes of experimentally recorded impulses in Fig.~\ref{idealpotential}. The dependence of oscillation period on excitability $\phi$ is linear for $\phi \in [0.01,0.05]$ and cubic for $\phi \in [0.05,0.07]$ (Fig.~\ref{Period_vs_Fi}).

How long can the oscillations last? In our experiments, the oscillations in a 50\ce{\mu}L LM lasted up to an hour. The amplitude decreases with time due to exhaustion of catalyst in the droplet, however the most typical cause of oscillations ceasing was breakage of the LMs. Generally, repeated cycles of freezing and warming caused disruption of the hydrophobic particle `skin' of a LM, resulting in the cargo being spilled. 

How can the observed phenomena be used in unconventional computing? As Horowitz and Hill mention in their famous ``The Art of Electronics'' --- ''A device without an oscillator either doesn't do anything or expects to be driven by something else (which probably contains an oscillator).''~\cite{horowitz1980art}. We produced a chemical analog of an electronic temperature sensitive oscillator: an oscillator circuit for sensing and indicating temperature by changing oscillator frequency with temperature~\cite{kleinberg1986temperature,blodgett1999temperature}. Future BZ computing devices will be hybrid chemical-electronic devices, needing components to generate wave-forms. The BZ LMs \emph{per se} are sources of (relatively) regular space pulses. We experimentally demonstrated that the frequency of the pulses can be switched from high to low by freezing the BZ LMs. This realisation could be used in future large-scale ensembles of BZ LMs which approximate fuzzy-logic many-argument functions, where inputs are represented by temperature gradients and outputs are dominating frequencies of the oscillations in the ensembles. 

\section*{Acknowledgement}

This research was supported by the EPSRC with grant EP/P016677/1 ``Computing with Liquid Marbles''.

\bibliographystyle{achemso3} %ACS (inc Langmuir) formatting
\bibliography{bibliography}
\end{document}